\documentstyle[preprint,aps]{revtex}
\input{epsf}
\begin{document}
\draft
%
%
%
\vspace{15cm}
\title{Evidence for $\nu_\mu \to \nu_e$ Oscillations \\
       from Pion Decay in Flight Neutrinos}
\vspace{5cm}
%
%
%
\author{
C.   Athanassopoulos,$^{11}$ 
L.B. Auerbach,$^{11}$ 
R.L. Burman,$^6$ 
D.O. Caldwell,$^3$ 
E.D. Church,$^1$\\
I.   Cohen,$^5$ 
B.D. Dieterle,$^9$ 
J.B. Donahue,$^6$ 
A.   Fazely,$^{10}$ 
F.J. Federspiel,$^6$ 
G.T. Garvey,$^6$\\
R.M. Gunasingha,$^7$ 
R.   Imlay,$^7$ 
K.   Johnston,$^8$ 
H.   J. Kim,$^7$ 
W.   C. Louis,$^6$ 
R.   Majkic,$^{11}$\\
J.   Margulies,$^{11}$ 
K.   McIlhany,$^{1}$ 
W.   Metcalf,$^7$ 
G.B. Mills,$^6$ 
R.A. Reeder,$^9$ 
V.   Sandberg,$^6$\\
D.   Smith,$^4$ 
I.   Stancu,$^1$ 
W.   Strossman,$^1$ 
R.   Tayloe,$^6$ 
G.J. VanDalen,$^1$ 
W.   Vernon,$^2$\\
N.   Wadia,$^7$ 
J.   Waltz,$^4$ 
D.H. White,$^6$ 
D.   Works,$^{11}$ 
Y.   Xiao,$^{11}$ 
S.   Yellin$^3$\\
(LSND Collaboration)\\
\mbox{}\\
}
\address{$^1$ University of California, Riverside, CA 92521}
\address{$^2$ University of California, San Diego, CA 92093}
\address{$^3$ University of California, Santa Barbara, CA 93106}
\address{$^4$ Embry Riddle Aeronautical University, Prescott, AZ 86301}
\address{$^5$ Linfield College, McMinnville, OR 97128}
\address{$^6$ Los Alamos National Laboratory, Los Alamos, NM 87545}
\address{$^7$ Louisiana State University, Baton Rouge, LA 70803}
\address{$^8$ Louisiana Tech University, Ruston, LA 71272}
\address{$^9$ University of New Mexico, Albuquerque, NM 87131}
\address{$^{10}$ Southern University, Baton Rouge, LA 70813}
\address{$^{11}$ Temple University, Philadelphia, PA 19122}
%
%
%
%
\date{\today}
\maketitle
\newpage
%
%
%
\begin{abstract}
A search for $\nu_\mu \to \nu_e$ oscillations has been conducted at the Los 
Alamos Meson Physics Facility using $\nu_\mu$ from $\pi^+$ decay in flight. 
An excess in the number of beam-related events from the 
$\nu_e \, C \to e^- \, X$ inclusive reaction is observed. 
The excess is too large to be explained by normal $\nu_e$ contamination in the 
beam at a confidence level greater than 99\%. 
If interpreted as an oscillation signal, the observed oscillation probability 
of $(2.6 \pm 1.0 \pm 0.5) \times 10^{-3}$ is consistent with the previously 
reported $\bar\nu_\mu \to \bar\nu_e$ oscillation evidence from LSND.
\end{abstract}
%
%
%
\section{Introduction}
\subsection{Motivation}

In this paper we describe a search for neutrino oscillations from pion
decay in flight (DIF). 
These data were obtained using the Liquid Scintillator Neutrino Detector (LSND) 
described in Ref.\cite{bigpaper1}. 
The result of a search for $\bar\nu_\mu \to \bar\nu_e$ oscillations, using 
a $\bar\nu_\mu$ flux from muon decay at rest (DAR), has already been reported 
in Ref.\cite{bigpaper2}, where an excess of events was interpreted as evidence 
for neutrino oscillations. 
The present paper provides details of an analysis of the complementary
process $\nu_\mu \to \nu_e$ from neutrinos generated from $\pi^+$ DIF. 

If indeed neutrino oscillations of the type $\bar\nu_\mu \to \bar\nu_e$ do 
occur, then $\nu_\mu \to \nu_e$ transitions must occur also. 
It is therefore important to search for the $\nu_\mu \to \nu_e$ transition 
to demostrate that the DAR signal is due to oscillations, instead of being a 
property of the $\mu^+$ decay. 
The $\pi^+$ DIF process provides a good setting for this search. 
It has completely different backgrounds and systematic errors from the DAR 
process, while providing an independent measurement of the same oscillation 
phenomena observed in the DAR measurement. 
Any excess of events in this analysis would support the neutrino oscillation
hypothesis.

The phenomenon of neutrino oscillations was first postulated by Pontecorvo 
\cite{bruno} in 1957. 
The underlying theory has been described in detail in standard textbooks. 
A general formalism for neutrino oscillations would involve 6 parameters 
describing the mixing of all three generations and the possibility of CP 
violation. 
In general, a $\nu_\mu$ beam can oscillate into both $\nu_e$ and $\nu_\tau$ 
with different amplitudes and different distance scales, set by the 
three-generation mixing angles and the three mass-squared differences. 
In the present case a relatively pure $\nu_\mu$ beam is produced at the
source. 
The LSND detector is sensitive to the $\nu_e$ state and thus, for simplicity, 
we approximate the process by a two-generation mixing model.
The oscillation probability can then be written as
\begin{equation}
P=\sin^2 2\theta \ \sin^2 \left(1.27 \ \Delta m^{2}{L\over E_\nu}\right),
\end{equation}
where $\theta$ is the mixing angle, $\Delta m^2$ (eV$^2$/c$^4$) is the 
difference of the squares of the masses of the appropriate mass eigenstates, 
$L$ (m) is the distance from neutrino production to detection, and $E_\nu$ 
(MeV) is the neutrino energy. 
The discussion is limited to this restricted formalism solely as a basis for 
experimental parameterization, and no judgement is made as to the simplicity
of the actual situation. 
\subsection{Comparison with other experiments}

In Ref.\cite{bigpaper2} the evidence restricting neutrino oscillation 
parameters is briefly reviewed. 
The salient features of that review are repeated here. 
There have been a series of experiments using beams derived from pion DIF 
which consist dominantly of $\nu_\mu$ with a small $\nu_e$ contamination.
The most sensitive experiment was at Brookhaven in a specifically designed 
long baseline oscillation experiment, E776 \cite{wonyong}.
This limit is shown in Figure \ref{fig:loglik} along with the favored region 
obtained by the LSND experiment. 
The limiting systematic error in E776 is a photon background from $\pi^{\circ}$ 
production, where one $\gamma$ is misidentified as an electron and the second 
$\gamma$ is not seen. 
The CCFR experiment \cite{ccfr} provides the most stringent limit on 
$\nu_\mu \to \nu_e$ oscillations near $\Delta m^2 \sim 350$ eV$^2$/c$^4$, but 
their limits are not as restrictive as E776 for values of $\Delta m^2 < 300$ 
eV$^2$/c$^4$. 
The KARMEN experiment \cite{karmen} has searched for $\nu_\mu \to \nu_e$ 
oscillations using neutrinos from pion DAR. 
These neutrinos are monoenergetic, and the signature for oscillations is  
an electron energy peak at about 12 MeV. 
This method has very different backgrounds and systematics compared
to the previous experiments but, unfortunately, does not yet have 
statistical precision sufficient to affect the exclusion region of 
Figure \ref{fig:loglik}. 
The KARMEN experiment also has searched for $\bar\nu_\mu \to \bar\nu_e$
oscillations and has produced the exclusion plot shown in Figure 
\ref{fig:loglik}. 
This is currently the most sensitive limit experiment in this channel. 
KARMEN is located 18 m from the neutrino source, compared with 30 m for LSND.
The experiments have sensitivities, therefore, that peak at different values 
of $\Delta m^2$.

The most sensitive experiment searching for $\bar\nu_e$ disappearance is 
Bugey \cite{bugey} using a power reactor which is a prolific source of 
$\bar\nu_e$. 
The detectors at Bugey observe both the positron from the primary neutrino 
interaction and the capture energy (4.8 MeV) from neutron absorption on 
$\mbox{}^6 Li$. 
The resulting limit is also shown in Figure \ref{fig:loglik}.

The most sensitive searches for $\nu_\mu$ disappearance have been conducted 
by the CDHS \cite{cdhs} and CCFR \cite{ccfr} experiments.
In each case two detectors are placed at different distances from the 
neutrino source, which is a DIF $\nu_\mu$ beam without focusing. 
The limits obtained by these experiments exclude the region with 
$\sin^2 2\theta > 0.08$ for values of $\Delta m^2$ typically above 1 
eV$^2$/c$^4$ and are not as restrictive as the limits set by the appearance 
experiments described above. 
Finally, the E531 Fermilab experiment \cite{reay} searched for the appearance 
of tau decays from charged-current interactions in a high energy neutrino beam. 
This $\nu_\mu \to \nu_\tau$ oscillation search excludes the region with 
$\sin^2 2\theta > 0.005$ for values of $\Delta m^2$ above approximately 10 
eV$^2$/c$^4$. 
Recently, the CHORUS and NOMAD experiments at CERN have reached limits close 
to that set by the E531 experiment with only a fraction of the data analyzed 
and should reach sensitivities of the order of $3 \times 10^{-4}$ in 
$\sin^2 2\theta$ in the near future.
\subsection{Experimental method}

LSND was designed to detect neutrinos originating in a proton target and 
beam stop at the Los Alamos Meson Physics Facility (LAMPF), and to search 
specifically for both $\bar\nu_\mu \to \bar\nu_e$ and $\nu_\mu \to \nu_e$ 
transitions with high sensitivity. 
This paper focuses on the second of these two complementary searches.
The neutrino source and detector are described in detail in 
Ref.\cite{bigpaper1}, with a summary in Section II of this paper. 
For the DIF experimental strategy to be successful, the neutrino source must 
be dominated by $\nu_\mu$, while producing relatively few $\nu_e$ by 
conventional means in the energy range of interest. 
The detector must be able to recognize $\nu_e$ interactions with precision 
and separate them from other backgrounds, many not related to the beam. 
The $\nu_e$ from conventional sources are small in number and are described 
in detail in Section VII.

LSND detects $\nu_e$ via the inclusive charged-current reaction 
$\nu_e \, C \to e^- \, X$.
The cross section for this process has been calculated in the continuum random 
phase approximation (CRPA) \cite{kolbe1}. 
This calculation successfully predicts the $\nu_e \, C \to e^- \, X$ 
cross section from the $\mu^+$ DAR $\nu_e$ flux as measured by the LSND 
\cite{lsnd_nuec}, KARMEN \cite{karmen_nuec}, and E225 \cite{e225} experiments. 
A similar calculation however predicts too large a cross section for the 
process $\nu_\mu \, C \to \mu^- \, X$ at higher energies. 
A discussion of the cross section uncertainties and comparisons to the data 
is presented in Section VIII. 
The final state electron energy can range from zero to the incident neutrino 
energy minus 17.3 MeV, which corresponds to the binding energy difference 
between the initial nucleus and the final state nucleus in the ground state.

The oscillation search analysis uses the following strategy. 
Beam-unrelated backgrounds induced by cosmic-ray interactions are removed as 
much as possible by requiring a positive identification of the electron from 
the $\nu_e \, C \to e^- \, X$ reaction in the tank. 
The remaining beam-unrelated background events in the sample are subtracted by 
using the data taken while the beam is off (beam-off sample) to determine the 
level of such background. 
Notice that the beam-off data is very well measured as LSND records 
approximately 13 times more data while the beam is off than while it is on. 
This procedure yields the number of excess events above cosmic background due 
to beam-induced neutrino processes. 
The remaining beam-related backgrounds are then subtracted to determine any 
excess above the expectation from conventional physics.
The number and energy distribution of the excess events are used to determine 
a confidence region in the $(\sin^2 2\theta,\Delta m^2)$ parameter space.

This paper describes two independent analyses that use independent
reconstruction techniques and different event selections. 
This has allowed cross checks on the software and selection criteria and 
has resulted in a more efficient final event selection. 
They shall be referred to as ``analysis A'' and ``analysis B'' throughout this 
paper.
\subsection{Outline of the paper}

We present a brief description of the neutrino source and detector system 
in Section II. 
Section III describes the initial data selection for the DIF analysis, while 
the reconstruction algorithm and particle identification parameters are 
discussed in Section IV. 
Section V describes the event selection and efficiencies for two independent 
analyses. 
Distributions of the data are shown in Section VI. 
Section VII contains an assessment of the beam-induced neutrino backgrounds. 
Fits to the data and an interpretation of the data in terms of neutrino 
oscillations are presented in Section VIII. 
The conclusions are summarized in Section IX.
%
%
%
\section{Neutrino Beam, Detector and Data Collection}
\subsection{The neutrino source}

This experiment was carried out at LAMPF 
\footnote{
The accelerator was operated under the name LAMPF until October 1995 
when the name was changed to LANSCE (Los Alamos Neutron Scattering Center).
} 
using 800 MeV protons from the linear accelerator. 
Pions were produced from 14772 Coulombs of proton beam at the primary beam stop 
over three years of operation between 1993 and 1995. 
There were 1787 Coulombs in 1993, 5904 Coulombs in 1994, and 7081 Coulombs 
in 1995. 
The fraction of the total DIF neutrino flux produced in each of the three 
years was 12\% in 1993, 42\% in 1994, and 46\% in 1995. 
The flux in 1995 was slightly reduced with respect to the Coulomb fraction due 
to variations in the target conditions, which are described below.
The duty ratio is defined to be the ratio of data collected with beam on to 
that with beam off. 
It averaged 0.070 for the entire data sample, and was 0.072, 0.078, and 
0.060 for the years 1993, 1994, and 1995, respectively.

A detailed description of the neutrino flux calculations from pion DIF in the 
LAMPF beam is given in Ref.\cite{burman}. 
A 1 mA beam of protons on the A1, A2 and A6 targets produces pions that are 
the source of the DIF neutrino beam \cite{bigpaper1}. 
The primary source of neutrinos consists of a 30-cm long water target (A6) 
surrounded by steel shielding and followed by a copper beam dump. 
It is located approximately 30 m from the center of the detector. 
About 3.4\% of the generated $\pi^+ $ decay in flight due to the open space 
between the water target and the beam stop, producing a $\nu_\mu$ flux with 
energies up to approximately 300 MeV. 
Most of the positive pions that decay come to rest prior to decaying. 
They then decay through the DAR sequence that produces the DAR neutrino fluxes 
via $\pi^+ \to \mu^+ \nu_\mu$ and $\mu^+ \to e^+ \nu_e \bar\nu_\mu$, 
where the $\nu_e$ and $\bar\nu_\mu$ have a maximum energy of 52.8 MeV. 
Of the $\mu^+$ that decay, approximately 0.001\% decay in flight and produce 
a small $\nu_e$ contamination of the $\nu_\mu$ beam. 
Another small contamination comes from the decay mode $\pi^+ \to e^+ \nu_e$ 
with a branching ratio of $1.24 \times 10^{-4}$. 
Together, these sources of $\nu_e$ constitute the major $\nu_e$-induced 
background for the DIF analysis, as discussed in Section VII. 
The negative chain starting with $\pi^-$ leads to a smaller contamination of 
the beam with $\bar\nu_e$, because the $\pi^-$ production cross section is 
suppressed by a factor of about eight relative to the $\pi^+$.

The two upstream carbon targets A1 and A2 were used to generate pion and muon 
beams for an experimental program in nuclear physics. 
They are located approximately 135 m and 110 m, respectively, from the 
center of the detector. 
The flux from each target depended on the thickness as well as the 
proton energy in the primary beam reaching the target. 
They were originally 3 cm and 4 cm thick, respectively, and degraded slowly 
during the operation of the accelerator. 
Their thickness was monitored regularly during the runs and incorporated in 
the beam flux simulation. 

The DIF neutrino flux varies approximately as $r^{-2}$ from the average
neutrino production point, where $r$ is the distance traveled by the neutrino. 
In addition, there is a significant angular dependence of the neutrino flux 
with respect to the direction of the incident proton beam. 
Thus, the DIF neutrino flux reaching the LSND apparatus has been calculated 
on a three-dimensional grid that covers uniformly the entire volume of the 
detector. 
The DIF fluxes at the detector center are illustrated in Figure 
\ref{fig:a126_flux94} for the positive decay chains only. 
Figure \ref{fig:a126_flux94}(a) shows the $\nu_\mu$ flux from 
$\pi^+ \to \mu^+ \nu_\mu$, while Figures \ref{fig:a126_flux94}(b) and 
\ref{fig:a126_flux94}(c) show the most significant $\nu_e$ background 
sources from $\pi^+ \to e^+ \nu_e$ and $\mu^+ \to e^+ \nu_e \bar\nu_\mu$, 
respectively. 
Notice that the $\nu_\mu$ contributions from the A1 and A2 targets are 
generally small compared to that from A6. 
However, for $\nu_\mu \to \nu_e$ oscillations with low $\Delta m^2$ the 
$\nu_\mu$ flux from the two upstream targets can have a significant effect.

The systematic error on the DIF flux is estimated to be 15\%. 
The calculated flux is confirmed within 15\% statistical error by the 
LSND measurement of the exclusive $\nu_\mu C \to \mu^- \, \mbox{}^{12}N_{gs}$ 
reaction \cite{lsnd_numuc}. 
This transition is very well understood theoretically, and the measurement 
is very clean due to the three-fold space-time correlations between the 
muon and the resulting decay electron and the positron emerging from the 
$\mbox{}^{12}N_{gs}$ $\beta$-decay. 
An independent beam flux simulation, based almost entirely on GEANT 3.21 
\cite{geant321}, has been developed in order to check the previous calculations 
and finds good agreement between calculated neutrino fluxes \cite{nim_beammc}.
\subsection{The detector and veto shield}

The detector consists of a steel tank filled with 167 metric tons of liquid 
scintillator and viewed by 1220 uniformly spaced $8 ''$ Hamamatsu 
photomultiplier tubes (PMT). 
The scintillator medium consists of mineral oil $(CH_2)$ with a small 
admixture (0.031 g/l) of butyl-PBD. 
This mixture allows the detection of both $\check{\rm C}$erenkov 
and isotropic scintillation light, so that the on-line reconstruction software 
provides robust particle identification (PID) for $e^\pm$, along with the 
event vertex and electron direction. 
The electronics and data acquisition (DAQ) systems were designed to detect 
related events separated in time.
This is necessary both for neutrino-induced reactions and for cosmic-ray 
backgrounds. 

Despite 2.0 kg/cm$^2$ shielding above the detector tunnel, there remains a 
very large background to the oscillation search due to cosmic rays, which is 
suppressed by about nine orders of magnitude to reach a sensitivity limited by 
the neutrino source itself. 
The 4 kHz cosmic-ray muon rate through the tank, of which about 10\% stop and 
decay in the scintillator, is reduced by a veto shield to a 2 Hz rate. 
The veto shield encloses the detector on all sides except the bottom. 
Additional counters were placed below the veto shield after the 1993 run to 
reduce cosmic-ray background entering through the bottom support structure.  
The main veto shield \cite{veto} consists of a 15-cm layer of liquid 
scintillator in an external tank, viewed by 292 uniformly spaced $5 ''$ EMI 
PMTs, and 15 cm of lead shot in an internal tank. 
This combination of active and passive shielding tags cosmic-ray muons that 
stop in the lead shot. 
The veto shield threshold is set to 6 PMT hits. 
Above this value a veto signal holds off the trigger for 15.2 $\mu$s while 
inducing an 18\% dead-time in the DAQ. 
A veto inefficiency $< 10^{-5}$ is achieved off-line with this detector for 
incident charged particles. 
The veto inefficiency is larger for incident cosmic-ray neutrons.
\subsection{Detector simulation}

A GEANT 3.15-based Monte Carlo is employed to simulate interactions in the LSND 
tank and the response of the detector system \cite{lsndmc}. 
It incorporates the important underlying physical processes such as energy loss 
by ionization, Bremsstrahlung, Compton scattering, pair production, and 
$\check {\rm C}$erenkov radiation. 
It also includes detector effects such as wavelength dependent light 
production, reflections, attenuation, pulse signal processing, and data 
acquisition. 
Much of the input to the detector response package was measured either in a 
test beam or in a controlled setting. 
Models for the transmission and absorption of light in the tank liquid were 
determined from measured data. 
The PMT characteristics, such as wavelength dependent quantum efficiencies, 
pulse shapes, and reflection characteristics, were measured and the results 
used in the simulation. 
The simulation is calibrated below 52.8 MeV using Michel electrons from the 
decay of cosmic-ray muons that stop in the detector volume. 
The properties of the scintillator, including absorption length and 
detailed characteristics of $\check{\rm C}$erenkov radiation in this medium, 
are all checked in this way. 
The extrapolation to higher energies is then made using the MC simulation, 
which correctly incorporates the behavior of electrons in the detector medium.

The primary Monte Carlo data set employed to calculate selection efficiencies 
is called the DIF-MC data set. 
The neutrino flux and energy spectrum were calculated at 25 points throughout 
the detector volume. 
The DIF-MC sample is created by folding the calculated $\nu_\mu$ energy 
spectrum with the cross section predicted by the CRPA model to generate 
$\nu_e \, C \to e^- \, X$ interactions throughout the tank. 
This corresponds to 100\% transmutation of the $\nu_\mu$ beam to $\nu_e$. 
The events were generated inside the surface formed by the faces of the PMTs.
%
%
%
\section{Initial Data Selection}
The signature for the DIF oscillation search is the presence of 
an isolated, high-energy electron ($60 < E_e < 200$ MeV) in the detector 
from the charged-current reaction $\nu_e \, C \to e^- \, X$. 
The lower energy cut at 60 MeV is chosen to be above 
the Michel electron energy endpoint of 52.8 MeV, while the upper energy cut 
at 200 MeV is the point where the beam-off background starts to increase 
rapidly and the signal becomes negligible. 
The analysis relies solely on electron PID in an energy regime for which 
no control sample is available. 
Furthermore, with the exception of the $\nu_e C$ reaction leading to the 
$^{12}N$ ground state, there are no additional correlations that help improve 
the detection of the signal. 

The PID parameters used in the DAR analysis, $\chi_a$, $\chi_r$, $\chi_t$ and 
$\chi_{tot}$ - as defined in Ref.\cite{bigpaper2} - have been used for the 
DIF analysis as an initial data selection. 
The disadvantages in this higher energy regime are that they do not 
discriminate adequately against a large beam-off background and are sensitive 
to energy extrapolation. 
Thus, the initial selection uses loose cuts based on the measured distributions 
of $\chi_r$, $\chi_a$ and $\chi_{tot}$ \cite{bigpaper2} just below the Michel 
energy endpoint ($50 < E_e < 52$ MeV), without any energy corrections. 
As demonstrated by MC simulations of the DIF data (DIF-MC), this selection is 
effective in identifying electron events from the 
$\nu_e \, C \to e^- \, X$ reaction. 
Over the energy interval of interest ($60 < E_e < 200$ MeV), the calculated 
efficiency is $98.1 \pm 1.7$\%.

In order to reduce the cosmic-ray muon induced background, we require the veto 
shield PMT hit multiplicity be $<$ 4 for the data sample. 
In addition, the events were required to be reconstructed within a fiducial 
volume that extends up to 35 cm from the PMT faces. 
Space-time correlations have been used in the initial selection to reduce the 
background generated by the cosmic-ray muons, either directly or through the 
decay Michel electron. 
These correlations are described in the following subsections. 
\subsection{Future correlations}

Despite the veto shield hit multiplicity requirement, some cosmic-ray muons 
contaminate the sample. 
This is seen in Figure \ref{fig:fdt9495}, which illustrates the distribution 
of the time difference between the current event and all the following ones, 
up to 51.2 $\mu$s. 
The fit to an exponential plus a constant reveals a time constant of 2.18 
$\mu$s, identifying stopped cosmic-ray muons.
Cosmic-ray muons that stop and decay in the detector are uniquely identified 
by the following Michel electron. 
As illustrated in Figure \ref{fig:prmuon}, there is a correlation between the 
muon tank hit multiplicity and the distance between the reconstructed vertices 
of the muon-electron pair. 
The difference in the samples shown in Figures \ref{fig:prmuon}(a) and 
\ref{fig:prmuon}(b) is briefly discussed below. 

Cosmic-ray muon events typically generate a high veto shield hit multiplicity. 
In order to suppress this high-rate background, as already mentioned in the 
previous section, the DAQ imposes a 15.2 $\mu$s dead-time after each event 
with a veto shield hit multiplicity $\ge$ 6. 
Furthermore, all events with high veto shield hit multiplicities get a simpler 
event vertex reconstruction than the one described in Ref.\cite{bigpaper1}, 
and no direction reconstruction. 
Correlations obtained from these data are shown in Figure \ref{fig:prmuon}(a). 
The number of cosmic-ray muons with veto shield hit multiplicities $<$ 6 is 
much lower than the ones with multiplicities $\ge$ 6, which explains the 
smaller size of the sample shown in Figure \ref{fig:prmuon}(b). 
Also, since these muons get both a full vertex and direction fit, the distance 
correlation between the muon-electron pair is tighter. 
In both cases, the distance correlation between the muon-electron pair 
degrades with increasing muon tank hit multiplicity (or equivalently, energy) 
due to the on-line reconstruction algorithm which always assumes a point-like 
event. 

All events that are followed in the next 30 $\mu$s by an event with a tank 
hit multiplicity between 200 and 700 (typical for Michel electrons) are 
possible candidates for stopped cosmic-ray muons. 
If, in addition, the current event has a tank hit multiplicity above 600 or is 
reconstructed closer than 200 cm to the following Michel electron candidate, 
the current event is eliminated.
The events that are eliminated by this selection are almost always followed by 
Michel electrons, as shown in Figure \ref{fig:future_te}.
This selection criterion is very powerful in rejecting cosmic-ray muons 
and has a high efficiency for keeping candidate electron events from the 
$\nu_e \, C \to e^- \, X$ reaction. 
This efficiency is calculated to be $99.6 \pm 0.4$\%.
\subsection{Past correlations}

Similarly, the time difference between the current event and all of the 
previous activities provides a distribution indicative of Michel electrons 
from stopped cosmic muons in the sample, as shown in Figure 
\ref{fig:pastdt}(a). 
Despite the energy requirement of at least 60 MeV there is still a small 
contamination from the tail of the Michel electron energy spectrum. 
Although this problem disappears at energies above 80 MeV, we choose to impose 
the following selection over the entire energy regime to maintain an energy 
independent selection efficiency. 
We require that the current event have no activities in the previous 
30 $\mu$s with a tank hit multiplicity above 600 or closer than 200 cm. 
After imposing this cut, the time distribution with respect to previous 
events becomes flat, as shown in Figure \ref{fig:pastdt}(b). 
Although this cut is powerful in rejecting the high-end tails of the Michel 
electron spectrum, this selection has an efficiency of only $85.5 \pm 0.5$\%, 
due to the fact that it covers the entire energy interval.
%
%
%
\section{Event Reconstruction and Electron Identification}
\subsection{Introduction}

The event reconstruction and PID techniques that are used in the DIF analysis 
were developed to utilize fully the capabilities of the LSND apparatus. 
The basis for the reconstruction is a simple single track event model, 
parametrized by the track starting position and time $(x,y,z,t)$, direction 
$(\varphi,\theta)$, energy $(E)$, and length $(l)$. 
The coordinate system used throughout this analysis is located at the 
geometrical center of the detector, with the $z$-axis along the cylindrical 
axis of the tank (approximately parallel to and along the beam direction) and 
the $y$-axis vertical, pointing upwards. 
The expected PMT photon intensity and arrival time distributions for any given 
event are calculated from these parameters and the result is compared with the 
measured values. 
A likelihood function that relates the measured PMT charge and time values to 
the calculated values is used to determine the best possible event parameters 
and at the same time provides PID.

As mentioned in the introduction, two independent sets of reconstruction 
software were developed as a cross check of the analysis results. 
The two algorithms follow similar overall strategies but differ in detail and 
implementation. 
The main differences lie in the parametrization of the various likelihoods and 
probability distributions that describe the detector response, and in the 
set of underlying event parameters used to describe the event. 

The electron identification is based on the relative likelihood of the 
measured PMT charges and times under the assumption that the source track is 
an electron. 
A detailed description of the physical processes in the tank can be found in
Ref. \cite{bigpaper1}.
Relativistic tracks in the detector generate light that falls into three 
categories: isotropic scintillation light that is directly proportional to the 
energy loss in the medium, direct $\check {\rm C}$erenkov light emitted in a 
$47^\circ$ cone about the track direction, and isotropic scattered 
$\check {\rm C}$erenkov light. 
These three components occur in roughly equal proportions (0.35:0.32:0.33) for 
relativistic particles. 
Only the isotropic scintillation component occurs for non-relativistic charged 
particles. 
This difference forms the basis for distinguishing electrons from 
non-relativistic particles such as neutrons and protons.

Each of the three light components has its own characteristic emission time 
distribution. 
The scintillation light has a small prompt peak plus a large tail which 
extends to hundreds of nanoseconds. 
The direct $\check {\rm C}$erenkov light is prompt and is 
measured with a resolution of approximately 1.5 ns. 
The scattered $\check {\rm C}$erenkov component has a time distribution 
between the direct $\check {\rm C}$erenkov light and the scintillation light, 
with a prompt peak and a tail that falls off more quickly than scintillation 
light.

The two reconstruction algorithms used in the DIF analysis are based on 
maximizing the charge and time likelihood on an event-by-event basis. 
For any given event defined by the set of parameters $\vec\alpha$,
\begin{equation}
\vec\alpha  = \left(x,y,z,t,\varphi,\theta,E,l\right),
\end{equation}
the event likelihood for measuring the set of PMT charges $(q_i)$ and times 
$(t_i)$ is written as a product over the 1220 individual tank PMTs as:
\begin{equation}
{\cal L}_{event} = \prod_{i=1}^{1220} {\cal L}_q(q_i;\vec\alpha) \,
                                      {\cal L}_t(t_i;\vec\alpha).
\end{equation}
Reversing the meaning of the likelihood function, ${\cal L}_{event}$ is the 
likelihood that the event is characterized by the set $\vec\alpha$, given 
the set of measured charges $(q_i)$ and times $(t_i)$. 
Maximizing the event likelihood ${\cal L}_{event}$ (or equivalently 
minimizing $- \ln {\cal L}_{event}$) with respect to $\vec\alpha$ determines 
the optimal set of event parameters.

The predicted likelihoods for PMT charges and photon arrival times are based 
on distributions measured from a large sample of Michel electrons from stopped 
cosmic-ray muon decays, as described below. 
Analysis A uses the entire spectrum of Michel electrons, whereas analysis B 
uses only electrons with 38 $<$ $E_e$ $<$ 42 MeV, henceforth referred to as 
``monoenergetic''. 
The upper edge of the Michel spectrum (52.8 MeV) is used to calibrate the 
energy scale of the system. 
The Michel electrons are well below the critical energy of 85 MeV and result 
in short track segments. 
The extension to longer, higher energy electron tracks is made by allowing for 
multiple discrete sources on the track. 
This is done either with two sources only along the track and fitting the 
distance between them (i.e., the track-length) in analysis A, or with a 
variable number of points, as determined from the energy of the event, 
distributed equidistantly along the track in analysis B. 
The energy dependence of the event likelihood is determined from the MC 
simulation. 
In the following subsections we briefly describe the charge and time 
likelihoods for a pointlike source of light.
\subsection{Charge likelihood}

Let us consider a pointlike light source located at $(x,y,z)$ in the detector 
and direction determined by $(\varphi,\theta)$ in spherical coordinates. 
For low energy relativistic electrons the track length is comparable with the 
dimensions of the PMTs, and thus the pointlike approximation provides a good 
model. 
The isotropic scintillation and scattered $\check{\rm C}$erenkov light have a 
combined strength $\Phi$ (photons per steradian), where $\Phi$ is proportional 
to the energy $E$ of the event. 
The strength of the anisotropic direct $\check{\rm C}$erenkov light is 
parametrized as $\rho \Phi$, while the angular dependence is given by a nearly 
Gaussian function, $f(\cos \theta_e)$. 
The angle $\theta_e$ is the angle with respect to the reconstructed event 
direction of the event and the function is normalized such that
\begin{equation}
\int_0^{\pi} f(\cos\theta_e) \sin\theta_e \, d\theta_e = 1.
\end{equation}
The function $f(\cos\theta_e)$, as determined from the data, is shown in 
Figure \ref{fig:cercosth_me_mc95}, with a vertical offset induced by the 
isotropic light. 

The average number of photoelectrons (PEs) $\mu$ expected at a phototube 
of quantum efficiency $\varepsilon$, at a distance $r$ from the source, and 
subtending a solid angle $\Omega$ is given by
\begin{equation}
\mu = \varepsilon \, \Phi \, F\left(\cos\theta_e,r \right) \, \Omega
\label{eq:mudefa}
\end{equation}
in analysis A. 
The function $F\left(\cos\theta_e,r \right)$ is determined directly from the 
Michel data. 
In analysis B $\mu$ is parametrized as
\begin{equation}
\mu = \varepsilon \, \Phi \, \left[ e^{-r/\lambda_s} \, \Omega_s + 
\rho \, f(\cos\theta_e) \, e^{-r/\lambda_c} \, \Omega_c \right].
\label{eq:mudefb}
\end{equation}
The parameters $\lambda_s$ and $\lambda_c$ are the attenuation lengths for 
scintillation and direct $\check{\rm C}$erenkov light in the tank liquid, 
respectively. 
While these are indeed expected to be different for the different components of 
the light, the effective solid angles subtended by the PMT, $\Omega_s$ and 
$\Omega_c$ should, in principle, be identical. 
The difference is induced by the finite size of the PMTs and by the difference 
in the angular distributions of the two light sources \cite{lsndtn106}. 
Both effective solid angles have been determined from the data. 
Notice that although the individual quantum efficiencies of the PMTs as well 
as the attenuation lengths are wave-length dependent, we use only global 
effective values as determined from the data.

The probability of measuring $n$ PEs in the presence of the light source 
is then given by a Poisson distribution of mean value $\mu$,
\begin{equation}
P(n;\mu) = \frac{1}{n!} \, e^{-\mu} \, \mu^n,
\end{equation}
with $\mu$ given by Eq.(\ref{eq:mudefa}) or Eq.(\ref{eq:mudefb}). 
However, since the LSND PMTs measure charge and not the number of PEs, the 
probability of measuring a charge $q$ for a predicted value $\mu$ is given by
\begin{equation}
{\cal P}(q;\mu) = \sum_{n=0}^{\infty} P(q;n) P(n;\mu),
\end{equation}
where the $P(q;n)$ functions are the charge response functions (CRFs) of the 
PMTs, i.e. the probability of measuring a charge $q$ given a number of PEs $n$. 
Since $\mu$ depends directly on the set of event parameters $\vec\alpha$, the 
probability ${\cal P}(q;\mu)$ determines directly the charge likelihood 
${\cal L}_q (q;\vec\alpha)$ for the PMT.

In analysis A the ${\cal P}(q;\mu$) functions are determined directly from the 
Michel data sample. 
In this sample, the predicted average number of PEs $\mu$ is calculated for 
every tube according to Eq.(\ref{eq:mudefa}), for all of the events. 
For PMTs in a given predicted $\mu$-bin, the distribution of the measured 
charge $q$, after proper normalization, yields directly the ${\cal P}(q;\mu)$ 
function required for the likelihood function above. 
The ${\cal P}(q;\mu)$ functions obtained in this way contain all instrumental 
effects incurred in measuring the charge, such as saturation and threshold 
effects.
Examples of these distributions are shown in Figure \ref{fig:chargelike} 
for two predicted charges, $\mu$ = 0.0-0.5 PE and $\mu$ = 2.5-3.0 PE.

Alternatively, analysis B obtains the two lowest CRFs, $P(q;0)$ and $P(q;1)$, 
and generates the higher $P(q;n)$ distributions as follows. 
The lowest CRF, $P(q;0)$, is just the Kronecker delta, $\delta_{q,0}$, since 
the probability of measuring a charge $q$ for no PEs vanishes identically 
for $q > 0$ and is unity when $q = 0$. 
The second CRF, $P(q;1)$, is the single-PE response of the PMTs, as illustrated 
in Figure \ref{fig:pq1_95}.
It is measured from low-intensity laser calibration data, taken during 
normal detector operation. 
The long tail of the single-PE charge distribution is probably due to 
collisions of electrons with material ahead of the first dynode. 
This effect is in good agreement with studies performed by the SNO experiment 
\cite{brodman}, which uses the same type of PMTs. 
The higher CRFs $(n \geq 2)$ are calculated by randomly sampling $P(q;1)$ $n$ 
times. 
With all CRFs normalized to unit area, the correct normalization of 
${\cal P}(q;\mu)$ is automatically insured,
\begin{equation}
\int_0^{\infty} {\cal P}(q;\mu) \, \mbox{d}q = 1, \,\,\,\, \forall \mu.
\end{equation}

Before going on to discussing the corrected time likelihood, we should point 
out that both reconstructions find a slightly longer attenuation length for the 
direct $\check {\rm C}$erenkov light in the MC Michel electrons than observed 
in the Michel electron data. 
This is needed to obtain agreement between the tank hit multiplicity and 
charge distributions obtained from the Michel electron data and the MC Michel 
electrons. 
This effect, which can be seen in Figure \ref{fig:cercosth_me_mc95}, has been 
shown not to affect significantly the charge and time likelihood distributions.
\subsection{Corrected time likelihood}

The corrected time $t_c$ of a PMT is defined as the measured PMT time after 
corrections for the fitted event vertex time and light travel time from the 
event to the PMT surface. 
For prompt light this peaks at $t_c = 0$ with an RMS of approximately 
1.5 ns. 
The time response functions for scintillation light and scattered 
$\check {\rm C}$erenkov light are more complicated and are determined from the 
Michel electron data. 
In addition, the time response functions depend upon the predicted charge. 
There is time slewing due to finite pulse rise time.
There is also time jitter from the distribution of transit time of electrons 
in the PMT for signals with small numbers of PEs.
Because the electronics responds to the first PE, above ten PEs the late tail 
in the distribution is negligible. 
Also, the amount of prompt $\check {\rm C}$erenkov light depends on whether 
or not the PMT is in the $\check {\rm C}$erenkov cone, as determined by 
$f(\cos\theta_e)$.

In analysis A the predicted average number of PEs $\mu$ is calculated for 
every PMT according to Eq.({\ref{eq:mudefa}), for all events in the sample. 
For PMTs in a given predicted $\mu$-bin, the distribution of the measured 
corrected time $t_c$, after normalization to unit area, provides directly 
the probability $P(t_c;\mu)$ required for the time likelihood function, 
${\cal L}_t(t;\vec\alpha)$. 
The $P(t_c;\mu)$s obtained in this way contain all instrumental effects 
incurred in measuring the time, such as time slewing and PMT jitter. 
Figure \ref{fig:timelike_a} shows two examples of these distributions for 
Michel electron data and MC simulated data. 
Both distributions are obtained in the ``isotropic'' region $\cos\theta_e < 
0.3$. 
Analysis A also measures the corrected time distributions in the 
$\check {\rm C}$erenkov ``peak'' region $(0.63 < \cos\theta_e < 0.73)$. 
An interpolation between these two distributions gives the time likelihood 
functions for the intermediate levels of direct $\check {\rm C}$erenkov light. 

The parametrization of the $P(t_c;\mu)$ distributions of analysis B is 
described next.
The timing distribution of the scintillation light for the LSND active medium 
has been measured to be of the form \cite{reeder}
\begin{equation}
f(t) = A_1 \, e^{-t/\tau_1} + \frac{A_2}{(1+t/\tau_2)^2} 
\hspace{5mm} \mbox{for} \hspace{5mm} t \, > \, 0.
\end{equation}
The two terms above represent the fast and the slow components of the light, 
with time constants $\tau_1$ = 1.65 ns and $\tau_2$ = 22.58 ns, respectively. 
The probability for observing a corrected time $t_c$ is
\begin{equation}
P(t_c) = \frac{1}{N} \int_0^{\infty} f(t') \, 
\exp \left[- \, \frac{1}{2\sigma^2} \, (t_c-t')^2\right] \, \mbox{d}t',
\end{equation}
which is the convolution of $f(t)$ with a time smearing function, assumed 
to be a Gaussian of width $\sigma$. 
The overall factor $1/N$ insures proper normalization to unity. 
While the integration of the fast component of the light can be analytically 
performed, this is not true for the slow component. 
Therefore we choose to parametrize the scintillation light as a superposition 
of three exponentially decaying functions,
\begin{equation}
f(t) = \sum_{i=1}^3 A_i(\mu) \, e^{-t/\tau_i(\mu)}.
\end{equation}
Both the amplitudes and the time constants are also allowed to vary as a 
function of the predicted amount of scintillation light $\mu$, as discussed 
above. 
The probability for recording a corrected time $t_c$ is thus
\begin{eqnarray}
P(t_c;\mu) & = & \frac{1}{N} \, \sum_{i=1}^3 A_i(\mu) \int_0^{\infty} \exp 
\left[
- \, \frac{1}{2\sigma^2(\mu)} \, (t_c-t')^2 - \frac{t'}{\tau_i(\mu)} \right] \, 
\mbox{d}t' \nonumber \\
& = & \frac{1}{N} \sum_{i=1}^3 A_i(\mu) 
\exp \left[\frac{\sigma^2(\mu)}{2\tau_i^2(\mu)}-\frac{t_c}{\tau_i(\mu)}\right] 
\mbox{Erfc}\left[\frac{1}{\sqrt{2}\sigma(\mu)}
\left(\frac{\sigma^2(\mu)}{\tau_i(\mu)}-t_c\right)\right],
\label{eq:ptc}
\end{eqnarray}
with the normalization factor $N$ given by
\begin{equation}
N = 2 \, \sum_{i=1}^3 A_i(\mu) \tau_i(\mu).
\end{equation}
Replacing $t_c$ by $t_c-t_0(\mu)$ in Eq.(\ref{eq:ptc}) above allows for 
additional time slewing corrections for the scintillation light. 
In LSND, the time slewing calibration is performed using laser calibration 
data \cite{nim_calib}, which is prompt. 
It is expected that the scintillation light should require additional 
time slewing corrections, which is confirmed by the data. 
A typical time probability distribution as a function of the corrected time 
is shown in Figure \ref{fig:tcbwdcer_95}(a). 
The quality of the fit to the data is excellent and shows that the 
parametrization given by Eq.(\ref{eq:ptc}) is a very good approximation. 

The time distributions for the direct $\check{\rm C}$erenkov light are also 
measured from the ``monoenergetic'' Michel electrons for different values of 
the predicted charge $\mu$. 
This is achieved by subtracting the appropriate underlying scintillation 
contributions from the corrected time distributions in the 
$\check{\rm C}$erenkov cone $(0.53 < \cos\theta_e < 0.79)$. 
The resulting distribution is shown in Figure \ref{fig:tcbwdcer_95}(b), which 
confirms the prompt character of the direct $\check{\rm C}$erenkov light.
\subsection{Electron identification}

The fitting procedures produce an accurate estimate of the amount of direct 
$\check{\rm C}$erenkov light in the event. 
In analysis A the level of $\check{\rm C}$erenkov light is determined after 
fitting the event to an electron model, i.e. a light source with an 
electron-like response for both charge and time likelihoods, that includes a 
full $\check{\rm C}$erenkov cone. 
With all other event parameters fixed, the amount of direct 
$\check{\rm C}$erenkov light is varied from none to the full amount for an 
electron event in order to maximize the event likelihood. 
This procedure determines a parameter $F_{Cer}$ which can thus 
vary between 0.0 (no direct $\check{\rm C}$erenkov light) and 1.0 (full 
amount of $\check{\rm C}$erenkov light).
In analysis B the amount of $\check{\rm C}$erenkov light is determined by 
varying all event parameters including $\rho$, the 
$\check{\rm C}$erenkov-to-scintillation density ratio. 
Figure \ref{fig:fcsr_me_mc95} shows the distribution of the 
$\check{\rm C}$erenkov-to-scintillation density ratio for ``monoenergetic'' 
Michel electrons, with a sharp peak at approximately $\rho = 0.5$. 
Moreover, for particles that are not expected to have any 
$\check{\rm C}$erenkov light (e.g. neutrons), the algorithm finds indeed a 
very low $\check{\rm C}$erenkov-to-scintillation fraction, which is still above 
zero because of fluctuations. 
This distribution is shown in Figure \ref{fig:neutrons_fcsr} for a sample of 
cosmic-ray neutrons in the same (electron-equivalent) energy range as the 
DIF sample. 
These have been tagged as neutrons by requiring the presence of a correlated 
$\gamma$ (from neutron capture on free protons, $n p \to d \gamma$) with a 
relatively high $R_\gamma$ parameter, as defined in Ref.\cite{bigpaper2}. 
Briefly, the $R_{\gamma}$ parameter is a quantity obtained from the $\gamma$ 
tank hit multiplicity, time and distance distributions with respect to the 
primary event. 
As shown in Ref.\cite{bigpaper2}, it provides an excellent tool for identifying 
correlated photons and rejecting the accidental ones.

The fitted optimum values of the charge and time negative log-likelihoods 
for the events are used as primary PID tools in the DIF analysis. 
In addition to the amount of $\check {\rm C}$erenkov light, they prove to be 
very different for non-electromagnetic events (e.g. neutrons) and provide very 
good discrimination against them, as will be shown in the next section. 
Analysis A makes use of the optimal values of the overall event charge and 
time likelihoods, ${\cal L}_q$ and ${\cal L}_t$, and the likelihoods calculated 
in the $\check{\rm C}$erenkov cone only $(0.53 < \cos\theta_e < 0.79)$, 
${\cal L}_{qc}$ and ${\cal L}_{tc}$. 
Analysis B uses instead the optimal values of the charge and time likelihoods 
obtained exclusively inside $(L_{Qcer},L_{Tcer})$ and outside 
$(L_{Qsci},L_{Tsci})$ the cone.

In both cases, the distribution of optimal likelihood values depend on 
the number of PMTs which were hit in a particular event. 
This is because the factor in the likelihood function for a PMT with a signal 
has a different functional form than for a tube without a signal. 
In order to remove this effect, the likelihoods are corrected as a function of 
the number of hit tubes. 
The mean value of the likelihood is then independent of the number of PMT hits. 
In addition, there is a dependence of the distribution of optimal likelihood 
values on the distance to the PMT wall, which is corrected for in an analogous 
way. 
Figure \ref{fig:michellike} shows the corrected distributions for analysis A as 
obtained from the entire Michel electron spectrum. 
Distributions from analysis B, as obtained from ``monoenergetic'' Michel 
electron events, are illustrated in Figure \ref{fig:llq_me_mc95}, before the 
hit multiplicity and distance corrections described above.

Both fitting algorithms significantly improve the position and direction 
accuracy over that used previously \cite{bigpaper2}.
The spatial position resolution is now approximately 11 cm and the angle 
resolution is approximately $6^{\circ}$ for electron events over the energy 
interval of interest for this analysis. 
The energy resolution is limited to 6.6\% at the Michel energy end-point, 
as stated in Ref.\cite{bigpaper1}. 
This is due to the width of the single-PE response of the PMTs (Figure 
\ref{fig:pq1_95}) and also due to tube to tube variations in the response 
functions.
%
%
%
\section{Data Selection and Efficiencies}
\subsection{Introduction}

The event selection presented in this section is designed to reduce 
cosmic-ray-induced background from the initial DIF sample. 
At the same time, the selection criteria must efficiently identify the 
electron of the final state in $\nu_e C \to e^- X$ interactions. 
These events have the following characteristics: 
they have little or no activity in the veto shield system; 
they are nearly uniformly distributed inside the detector; 
they have no excess activity either before or after the event time; 
and they yield a track inside the tank which is consistent with an electron, 
as identified through the characteristic scintillation and 
$\check {\rm C}$erenkov light. 
These are the only features available for electron event selection except in 
the rare case of a transition to the $^{12}N$ ground state, which subsequently 
$\beta$-decays with a 15.9 ms lifetime.

The beam-off background data and simulated DIF-MC electron events 
are used in order to choose the optimal selection value for each quantity in
a way unbiased by actual beam-on data. 
The sensitivity, or ``merit'' for the value of a selection parameter is defined 
as the efficiency $\epsilon$ (determined from the DIF-MC) divided by 
the square root of the number of selected beam-off events, $N_{off}$, 
scaled by the duty ratio $f$:
\begin{equation}
M \equiv \frac{\epsilon}{\sqrt{f\,N_{off}}}.
\end{equation}
Each selection parameter value is varied and the point of maximum sensitivity, 
or ``merit'', determines the optimal value of the selection. 
This method is independent of the beam-on data. 
Throughout this section the selection criteria for analyses A and B 
are discussed in parallel.
\subsection{Analyses A and B}

The cosmic-ray backgrounds are dominated by several types of processes. 
The level of all cosmic-ray-induced processes is measured in the beam-off 
data sample, and then the appropriate amount is subtracted from the final 
beam-on sample.

Cosmic-ray neutrons generate a non-electromagnetic background. 
A fraction of these neutrons evade the veto shield and enter the detector 
volume to interact with carbon nuclei and protons in the liquid. 
The interaction length is approximately 75 cm. 
Their presence in the DIF sample is due to the very loose initial electron 
identification selection and can be consistently demonstrated in three 
different ways, as follows:

The typical signature of neutron events is the 2.2 MeV correlated $\gamma$ 
that results from capture on free protons, $n p \to d \gamma$. 
These $\gamma$ candidates are recorded in a 1000 $\mu$s interval after the 
primary trigger. 
The time difference between the primary events of the entire DIF sample 
(beam on + off) and all subsequent $\gamma$ events is shown in Figure 
\ref{fig:gdt_all9495}. 
The fitted lifetime of $186.2 \pm 0.4$ $\mu$s, is in very good agreement 
with the known neutron capture time of 186 $\mu$s. 
The constant part of the fit determines the total number of accidental photons.
After subtracting this number from the total number of $\gamma$ candidates in 
the sample one obtains that {\em on average} every DIF event has one correlated 
photon. 
This is consistent with a considerable contamination of the DIF sample by 
cosmic-ray neutrons, since these are expected to have on average more than 
one associated $\gamma$.

Secondly, as shown in Figure \ref{fig:neutrons_fcsr}, neutron events do not 
produce significant $\check {\rm C}$erenkov light in the liquid. 
The distribution of the $\check {\rm C}$erenkov-to-scintillation density ratio,
$\rho$, for the entire DIF sample (beam on + off) is shown in 
Figure \ref{fig:fcsr_difall} for analysis B. 
The same distribution for cosmic-ray neutron events with similar deposited 
energy and for the DIF-MC sample are superimposed. 
This illustrates that most of the (beam-off) background is dominated by 
non-electromagnetic events, consistent with neutrons. 
In order to select electron-like events, analysis A requires $F_{Cer} > 0.7$ 
and analysis B requires $\rho > 0.4$, as dictated by the maximum merit 
algorithm. 
Figure \ref{fig:fcerrho_difmc} shows the $F_{Cer}$ and $\rho$ distributions 
for the DIF sample and the DIF-MC electron events after all other selection 
criteria have been applied.

Finally, the event charge and time likelihood parameters defined in 
Section IV are different for neutron and electron events. 
These charge and time likelihood parameters are used in differentiating 
electromagnetic particles that produce $\check {\rm C}$erenkov light from 
non-electromagnetic backgrounds. 
Both analyses rely on this identification, using slightly different criteria.

Analysis A uses a likelihood ratio, $LR_{event}$, defined by forming the 
product of the charge and time likelihoods in each of the regions for the DIF 
beam-off sample and dividing it by the same product for the DIF-MC electrons:
\begin{equation}
LR_{event} \ = \ 
\frac{{\cal P}(-\ln {\cal L}_q^{off})    \times 
      {\cal P}(-\ln {\cal L}_{qc}^{off}) \times 
      {\cal P}(-\ln {\cal L}_t^{off})    \times 
      {\cal P}(-\ln {\cal L}_{tc}^{off})}
     {{\cal P}(-\ln {\cal L}_q^{mc})     \times 
      {\cal P}(-\ln {\cal L}_{qc}^{mc})  \times 
      {\cal P}(-\ln {\cal L}_t^{mc})     \times 
      {\cal P}(-\ln {\cal L}_{tc}^{mc})}.
\end{equation}
Figure \ref{fig:pl_ll_dif_he_p} shows the individual distributions of the four 
${\cal P}(-\ln {\cal L}_x)$ functions for the beam-off data and for the DIF-MC 
electron data.
The ratio $LR_{event}$ tends to be large for events that are like cosmic-ray 
background and small for electron-like events. 
Electron events are identified by requiring $LR_{event}$ $<$ 0.5. 
The event time likelihood in the $\check {\rm C}$erenkov cone, $L_{tc}$ is 
also used in the selection, being sensitive to the presence of 
$\check {\rm C}$erenkov light. 
This parameter is required to have a value $<$ 1.1.
Analysis B uses only the individual time likelihoods for identifying 
electromagnetic particles, by requiring $L_{Tsci} < 1.15$ and $L_{Tcer} < 1.5$. 
The charge and time negative log-likelihoods are illustrated in 
Figure \ref{fig:llq_difall} for the entire DIF data (beam on + off), DIF-MC 
electron events and cosmic-ray neutron events. 

The second class of backgrounds is electromagnetic, and leads to events that 
are difficult to distinguish from pure electron events. 
Charged particles occasionally evade the veto shield and enter the liquid
volume. 
These cannot travel into the liquid very far without depositing large amounts 
of energy and are reconstructed with a position very close to the tank wall. 
Their reconstructed direction points predominantly into the detector and 
the track can be extrapolated back to the tank wall where the veto shield 
information can be used. 
The veto counter system that surrounds the detector provides PMT signals 
which are read out and recorded as are the tank PMTs. 
For events with a non-zero veto shield hit multiplicity, $N_{veto} > 0$, 
the reconstructed tracks in the detector are extrapolated backwards to 
intersect the veto shield. 
A corrected time difference, $t_{veto}$, between the veto shield hit closest 
in time and space and the extrapolated event time is defined. 
Selecting events with $|t_{veto}| > 50$ ns (analysis A) or $> 70$ ns 
(analysis B) discriminates against any cosmic-ray-induced activity around the 
detector near the event in question. 
The $t_{veto}$ distribution for the entire DIF data sample (beam on+off) is 
shown in Figure \ref{fig:tv_timing}. 
In addition, a direct cut on the total veto hit multiplicity, $N_{veto} < 3$, 
is required in analysis A. 
Analysis B does not impose this cut. 
Figure \ref{fig:vetoh_dif} shows the veto shield hit multiplicity distributions 
for the beam-on, beam-off and beam-excess events after all selection criteria 
of analysis B have been applied. 
The distribution for the beam-excess events is consistent with the veto shield 
accidental hit distribution.

High energy $\gamma$ rays, from $\pi^0$ produced by neutron interactions in 
the lead shielding of the veto shield, enter the detector fiducial volume 
without leaving a veto signal. 
Energy is deposited through Compton scattering or by pair conversion. 
The latter process dominates above the 85 MeV critical energy of the liquid. 
The $\gamma$ attenuation length is roughly 50 cm, the radiation length in the 
liquid. 
The charged particles resulting from their interactions in the liquid point 
into the detector volume. 
These events are difficult to distinguish from electrons of the 
$\nu_e \, C \to e^- \, X$ reaction in this detector on the basis of electron 
identification alone.

This class of backgrounds is characterized by its typical distribution of the 
length of the flight-path inside the detector. 
This quantity is defined as the distance between the reconstructed event vertex 
and the intersection of the backwards extrapolation along the reconstructed 
event direction with the PMT surface, $\Delta$, (analysis A) or with the tank 
steel wall, $S$ (analysis B). 
Although slightly different in their definitions, both quantities correspond 
to the distance a neutral particle would have to travel in the liquid before 
it interacts. 
Events were required to satisfy $\Delta > 175$ cm in analysis A and 
$S > 225$ cm in analysis B. 
The distributions of the $\Delta$ and $S$ variables are shown in Figure 
\ref{fig:deltas_difmc} for the DIF sample and the DIF-MC electron events after 
all other selection criteria have been applied.

Another possible source of cosmic-ray backgrounds that is reduced by the 
$\Delta$ and $S$ selections arises from $K_L^{\circ}$ decay 
inside the detector volume. 
The $K_L^{\circ}$ can travel into the detector where the $K_{e3}$ decay 
produces a positron (indistinguishable from an electron) and a $\pi^-$ 
($K_L^{\circ} \to \pi^- \, e^+ \, \nu_e$). 
The $\pi^-$ will stop and be absorbed, while the positron is in the energy 
range of interest. 
The other decay chain ($K_L^{\circ} \to \pi^+ \, e^- \, \bar\nu_e$) cannot 
contribute to the final DIF sample due to a degraded PID generated by the 
muon from the pion decay being virtually simultaneous with the electron, and 
due to space-time correlations with the positron from the subsequent muon 
decay.

The electron from the $\nu_e \, C \to e^- \, X$ reaction is backward peaked,
opposite the direction of the incident neutrino. 
Due to the geometry of the detector shielding, beam-off data favor the 
neutrino direction. 
Furthermore, the electron from one of the beam-induced backgrounds, the 
$\nu e \to \nu e$ elastic scattering, is also strongly peaked in the forward 
direction. 
The cosine of the angle between the reconstructed event direction and the 
incident neutrino direction, $\cos \theta_\nu$, is used to remove most of this 
background, by requiring $\cos \theta_\nu$ $<$ 0.8 in both analyses.

The veto system is very effective at rejecting cosmic-ray-induced backgrounds, 
but there were several penetrations in the system. 
A penetration at the lower upstream end of the veto system allows cables to 
enter the tank. 
For part of the data taking period there were several poorly performing PMTs 
at the top of the veto system. 
These regions were removed from the final data set of analysis A by requiring 
that the projected {\em entry} points of the events not lie in the regions
$(240^{\circ} < \phi_{xz} < 300^{\circ},-0.2 < \cos\theta_y<-0.6)$ and 
$(240^{\circ} < \phi_{xz} < 300^{\circ}, 0.6 < \cos\theta_y<1.0)$. 
The angles $\theta_y$ and $\phi_{xz}$ are defined with respect to the 
coordinate system of the detector defined at the beginning of Section IV.

The quantities 
$\Delta$, 
$F_{Cer}$,
$LR_{event}$, 
$L_{tc}$, 
$\cos\theta_\nu$, 
$t_{veto}$, 
$N_{veto}$ and 
the ``veto hot-spots'' 
(analysis A) and
$S$, 
$\rho$, 
$L_{Tcer}$, 
$L_{Tsci}$, 
$\cos\theta_\nu$ and 
$t_{veto}$ 
(analysis B) 
are used to select a final sample of events for the DIF analysis. 
The values of these selection criteria, efficiencies for $\nu_\mu \to \nu_e$ 
events, and rejection power are shown in the 
Tables \ref{tab:selection_a} and \ref{tab:selection_b} for analyses A and B, 
respectively. 
Event selection efficiency is defined with respect to events generated inside
the fiducial volume of the detector that extends all the way to the PMT 
surfaces ($d > 0$ cm).

Table \ref{tab:events} shows the number of beam-on, beam-off, and excess events
that result from the event selections described above. 
There is a clear excess of events above beam-unrelated backgrounds that is 
consistent with $\nu_e \, C \to e^- \, X$ reactions.

Average beam-unrelated backgrounds are determined by the total number of 
beam-off events times the beam duty ratio. 
A better estimate of the beam-unrelated background in the beam-on sample 
relies on further information contained in the beam-off event sample. 
The characteristic shape of the event direction distributions in the tank 
coordinate system for $\nu_e \, C \to e^- \, X$ signal events are different 
than those for the beam-off sample. 
Figure \ref{fig:cosy_phi} shows the distributions in $\cos\theta_y$ and 
$\phi_{xz}$ of the event directions for the final beam-off DIF sample as 
compared to the same distributions as obtained for the DIF-MC electrons.

It is possible to introduce a small bias into the event selection by using
the maximum ``merit'' or sensitivity method to select values for the 
selection criteria. 
Because the algorithm uses the beam-off data, it can pick points where the 
beam-off data has fluctuated down. 
Even though this should be a negligible effect, the level of beam-unrelated 
backgrounds in the beam-on sample can also be determined nearly independently 
of the number in the beam-off sample. 
This is done by performing a maximum likelihood fit to obtain the number of 
beam-unrelated background events in the beam-on sample. 
The two-dimensional $(\cos \theta_y,\phi_{xz})$ distribution of the beam-on 
event directions is fitted to a sum of the shapes expected for signal events 
and beam-unrelated background events. 
The likelihood for the total number of beam-unrelated events is weighted by 
the Poisson probability expected from the predicted beam-unrelated average. 
The results of this procedure are shown in Table \ref{tab:fitback} along 
with the results of using the product of the duty ratio and the number of 
beam-off events for each analysis. 
The probability that the number of observed beam-on events is a fluctuation is 
also shown in Table \ref{tab:fitback}. 
A systematic uncertainty of 21\% 
in the cross section, flux, and efficiency is included in the calculation.

The comparison of selections A and B along with the logical AND and logical OR 
of the two samples is shown in Table \ref{tab:andorisgm}. 
The number of beam-on events, background events, efficiencies, and resulting 
oscillation probabilities are all consistent within the statistical errors in 
the samples. 
Since the two analyses have low efficiencies, different reconstruction software, 
and different selection criteria, the overlap need not be large. 
The AND sample contains 8 beam-on events, which is consistent with the 11.6 
events expected by comparing the overlap of DIF-MC data and beam-off data. 
The final event sample is obtained as the logical OR of the events from 
analysis A and analysis B. 
This procedure minimizes the sensitivity of the measurement to uncertainties 
in the efficiency calculations and also yields a larger efficiency than the 
individual analyses.
The AND sample has the lowest efficiency for DIF electron events
and therefore the least sensitivity. 
The OR sample has the largest efficiency and hence the largest 
sensitivity to oscillation signals.
The probability that the backgrounds in the AND and OR samples fluctuate upward 
to the observed beam-on numbers are 0.18 and $1.1\times10^{-3}$ respectively.  
%
%
%
\section{Data Signal}
\subsection{Distributions of data}

Extensive checks have been performed on the final DIF OR sample to study 
the consistency with electron events from the $\nu_e \, C \to e^- \, X$ 
reaction. 
The distribution of $\cos \theta_\nu$, the cosine of the angle between the 
reconstructed electron direction and the incident neutrino direction, is 
shown in Figure \ref{fig:dif_uzxcs_or}. 
This distribution is slightly backwards peaked, as indeed expected from the 
$\nu_e \, C \to e^- \, X$ reaction, and agrees well with that obtained 
from the DIF-MC. 
The distance to the PMT surfaces for the final beam-excess data set is 
shown in Figure \ref{fig:dif_distxcs_or}, which also agrees with the 
DIF-MC expected distribution. 
Notice that the apparent depletion of events in the outer region of the 
fiducial volume is caused primarily by the $\Delta$ and $S$ selections of 
the two analyses. 
Small deviations from the original (on-line) reconstructed event vertices, 
induced by the new reconstruction algorithms, contribute to a smaller extent 
to this effect. 
The $x$, $y$ and $z$ distributions for the final beam-excess DIF sample 
are shown in Figure \ref{fig:dif_xyzxcs_or} and are in very good agreement 
with those obtained from the DIF-MC simulations. 
The distributions of the 40 beam-on events and 175 beam-off events in the 
$(x,y)$ and $(y,z)$ planes are illustrated in Figure 
\ref{fig:dif_xyyzonoff_or}. 
The energy distribution of the beam-excess events is illustrated in 
Figure \ref{fig:dif_excs_or}, together with the energy distribution of the 
beam-induced background and that expected from a positive $\nu_\mu \to \nu_e$ 
oscillation signal for large values of $\Delta m^2$. 
\subsection{Associated photons from neutron capture}

The $\nu_e \, C \to e^- \, X$ reaction is not normally expected to produce 
free neutrons at these energies. 
Only rarely (approximately 10\%) is a neutron knocked out by the incident 
neutrino, and then it is identified by the presence of a correlated 2.2 MeV 
$\gamma$ from the capture on free protons. 
Also, the small $\bar\nu_e$ contamination of the beam produces a small number 
of events with a correlated $\gamma$ via the inverse $\beta$-decay reaction 
$\bar\nu_e \, p \to e^+ \, n$. 
The correlated $\gamma$ identification relies on the $R_\gamma$ parameter 
mentioned earlier in the text, which in turn relies on the $\gamma$ tank hit 
multiplicity, time and distance distributions with respect to the primary 
event. 
The reconstruction algorithms used in the current analysis provide a better 
position resolution not only for the primary events, but also for the $\gamma$, 
as shown in Figure \ref{fig:gdr_oldnew}.
Using the sharper distance distribution between the $\gamma$ and the primary 
events in the calculation of $R_{\gamma}$ provides much better discrimination 
between correlated and accidental $\gamma$.

The distributions for the number of photons with $R_\gamma$ $>$ 1 are 
illustrated in Figure \ref{fig:corgammas_dif} for the final DIF beam-on, 
beam-off and beam-excess samples of analysis B. 
This particular value of the $R_{\gamma}$ cut accepts over 95\% of the 
correlated photons, while at the same time rejecting approximately 95\% of 
the accidental ones. 
The beam-induced excess yields a fraction of events with ``correlated'' 
photons $(R_{\gamma} > 1)$ that is consistent with that 
measured in the $\nu_\mu \, C \to \mu^- \, X$ channel, as reported in 
\cite{lsnd_numuc}.
\subsection{The transition to the $^{12}N_{gs}$}

The transition $\nu_e \, C \to e^- \, ^{12}N_{gs}$, which is expected to 
occur roughly 5\% of the time, is a useful signature in the search for 
$\nu_\mu \to \nu_e$ oscillations. 
It is nearly free of cosmic-ray background due to the detection of the 
space-time correlated positron from the $^{12}N_{gs}$ $\beta$-decay 
\cite{lsndtn108}. 
It is noteworthy that the ground state of $^{12}N$ has been extensively 
studied in (p,n) reactions \cite{n12gs}, as well as its analog, 
$\mbox{}^{12}C$ (15.11 MeV), in (e,e') and (p,p') reactions. 
The transition is well known and can be characterized successfully. 
The positron has an end-point energy of 17.3 MeV and a decay time constant of 
15.9 ms. 
The positron selection criteria are:
(i)   0.052 ms $<$ $\Delta t$ $<$ 45 ms; 
(ii)  reconstructed distance to the primary electron $<$ 100 cm; 
(iii) tank hit multiplicity $>$ 75 (in order to be above the accidental 
      $\gamma$ ray background); 
(iv)  positron energy $<$ 18 MeV; 
(v)   veto shield hit multiplicity $<$ 4. 
Using the same selection criteria for the primary electron as described above,
and in addition imposing the positron space-time correlations,
2 beam-on events and 1 beam-off event are observed. 
However, one of the two beam-on events has three correlated $\gamma$ and 
is thus not consistent with the ground state hypothesis. 
Eliminating this event from the sample, one obtains a beam-induced excess 
of $0.9 \pm 1.0$ events which is consistent with the expected 0.3-1.0 events, 
depending on the values of the oscillation parameters, as obtained from the 
inclusive analysis.
%
%
%
\section{Beam-Related Backgrounds}
This section discusses the beam-related backgrounds induced by neutrino 
interactions, which are not accounted for by the beam-off subtraction. 
Each of the backgrounds depends upon neutrino fluxes and cross sections that 
are energy dependent. 
The proper variation of efficiency with energy in analyses A and B is 
used in Sections V and VIII. 
In this section a generic, energy independent efficiency for electron 
events of 10\% is assumed for the sake of clarity, a number close to the 
actual average values for the two analyses. 
The effects of systematic uncertainties on the beam-related backgrounds are
crucial to the analyses. 
The major effects are discussed extensively in Section VIII.

There are four significant neutrino backgrounds in the $\nu_\mu \to \nu_e$ 
DIF oscillation search considered here. 
These backgrounds are: $\mu^+ \to e^+ \nu_e \bar\nu_\mu$ and 
$\pi^+ \to e^+ \nu_e$ DIF followed by $\nu_e C \to e^- X$ scattering; 
$\pi^+ \to \mu^+ \nu_\mu$ DIF followed by $\nu_\mu e \to \nu_\mu e$ 
elastic scattering; and $\pi^+ \to \mu^+ \nu_\mu$ DIF followed by 
$\nu_\mu C \to \nu_\mu C \pi^{\circ}$ coherent scattering. 
Backgrounds from $\nu_\mu C \to \mu^- X$ reactions are negligible. 
Muons that stop in the tank either decay or capture on carbon nuclei. 
The correlations in time and position between the muon and the secondary event 
removes them.
In the rare case that the second event is missed, the lone muon fails the 
electron identification, and the event is almost always below the 60 MeV 
electron equivalent energy limit. 
In the case of $\mu^-$ decay in flight, the long lifetime of the muon and the 
electron identification requirements reduce this background to a negligible 
level.

The significant backgrounds are summarized in Table \ref{tab:backgrounds} 
for reconstructed event energies between 60 MeV and 200 MeV. 
The volume used for normalization throughout this section is the $d > 0$ 
fiducial volume, as described earlier, which contains the equivalent
of $5.4 \times 10^{30}$ $CH_2$ molecules (or $4.3 \times 10^{31}$ electrons). 
For the combined 1993, 1994, and 1995 running periods there were 
$9.2 \times 10^{22}$ protons on target (POT).

The largest background is due to $\nu_e$ that come from $\mu^+$ DIF in the 
beam stop, followed by $\nu_e C \to e^- X$ scattering in the detector. 
This cross section is calculated in the CRPA model \cite{kolbe1}, as already 
mentioned. 
This results in a background of 3.8 events for the assumed efficiency. 
The next largest is background is due to $\pi^+ \to e^+ \nu_e$ DIF in the beam 
stop, followed by $\nu_e C \to e^- X$ scattering in the detector. 
The estimated background contribution is 1.9 events. 
The systematic error on this contribution is discussed in Section VIII. 
The previous two background catagories are produced by the same reaction as 
the DIF oscillation signal and are nearly impossible to distinguish from them 
on an event-by-event basis.

Another background is coherent $\pi^{\circ}$ production via the reaction 
$\nu_\mu C \to \nu_\mu C \pi^{\circ}$. 
This cross section has been calculated in Ref.\cite{reinsehgal}. 
Energetic electrons can be produced by the photons from the $\pi^{\circ}$ 
decay, which convert in the tank liquid and fake an electron. 
The fraction of $\pi^{\circ}$ that satisfy all selection criteria and are 
misidentified as electrons is 0.6. 
The estimated background contribution from coherent $\pi^{\circ}$ production 
is 0.3 events. 
Note that the non-coherent $\pi^{\circ}$ production is negligible at these 
energies. 

The last background considered is $\nu_\mu e \to \nu_\mu e$ elastic scattering 
on the $4.3 \times 10^{31}$ electrons in the $d > 0$ fiducial volume. 
This purely leptonic cross section is well known theoretically. 
The reaction is identified by the electron direction being nearly parallel to 
the incident neutrino direction. 
The fraction of electrons within the event selection region of 
$\cos \theta_\nu < 0.8$ is 0.05. 
The background contribution from $\nu_\mu e \to \nu_\mu e$ elastic scattering 
is estimated to be 0.1 events.

Table \ref{tab:backgrounds} shows the background estimates, where the total 
background is calculated to be $5.8 \pm 1.0$ events. 
The number of events expected for 100\% $\nu_\mu \to \nu_e$ transmutation 
is 4470 events.
%
%
%
\section{Interpretation of the Data}
This section describes the interpretation of the observed event excess in 
terms of the theoretically expected processes and a neutrino oscillation model.
The oscillation model employed here assumes two-generation mixing, as discussed 
in Section I. 
The confidence regions in the $(\sin^2 2\theta,\Delta m^2)$ parameter plane are 
calculated in this context. 
The effects of systematic errors are critical to the interpretation, and are 
described next.
\subsection{Systematic uncertainties}

The systematic uncertainties in this measurement arise from several sources. 
The dominant uncertainty comes from the knowledge of the underlying neutrino 
cross sections and the neutrino flux through the detector. 
The electron selection efficiency calculation also introduces some uncertainty.

The oscillation search relies on the knowledge of the 
$\nu_e \, C \to e^- \, X$ cross section in the 60-200 MeV electron energy 
range. 
The inclusive reaction has been calculated in the CRPA model \cite{kolbe1} and 
has been measured \cite{lsnd_nuec}-\cite{e225} by using the $\nu_e$ flux 
from $\mu^+$ DAR. 
The DAR flux is in turn measured by the well understood ground state 
$\nu_e \, C \to e^- \, \mbox{}^{12}N_{gs}$ inverse $\beta$-decay reaction. 
Both the ground state and the inclusive $\nu_e \, C$ reaction measurements 
agree well with the theoretical predictions, which indicates that both the flux 
and the cross section are predicted well. 
An extrapolation of the theoretical cross section must be made into the DIF 
energy region. 
We assign a systematic error of 10\% to this extrapolation, based upon the 
agreement between the measured $\nu_e \, C$ data and the theoretical CRPA 
prediction.

The DAR $\nu_e$ flux endpoint is at 52.8 MeV, below the region of interest for 
the DIF oscillation search. 
The DIF neutrino flux comes from pions that decay in flight rather than from 
stopped $\mu^+$. 
The $\nu_\mu \, C$ ground state and inclusive measurements of LSND 
\cite{lsnd_numuc} provide a check on the DIF flux. 
The $\nu_\mu \, C \to \mu^- \, \mbox{}^{12}N_{gs}$ ground state cross section 
is also well understood and is nearly independent of energy above its 123 MeV 
threshold. 
Thus, LSND measures the integral of the $\nu_\mu$ flux above threshold. 
The agreement between the predicted flux and the measured flux gives a 
constraint on the flux above threshold with an error of 15\%.
The $\nu_\mu \, C$ inclusive reaction cross section has a much stronger energy 
dependence than the ground state reaction. 
LSND measured this cross section \cite{lsnd_numuc} with high statistics and 
obtained a value that is approximately 45\% lower than the theoretically 
predicted value. 
The calculated {\em flux} $\times$ {\em cross} {\em section} does not agree 
with the measured data in this case. 
These neutrinos are in an energy range that overlaps with the DIF 
$\nu_{\mu} \to \nu_e$ energy range and represent the same $\nu_\mu$ flux that 
the DIF oscillation search uses. 
It is possible that the {\em flux} $\times$ {\em cross} {\em section} for the 
$\nu_e \, C \to e^- \, X$ reaction also follows this trend and is lower than 
what we have assumed. 
The consequences of this are discussed below.

The next important systematic error is the extrapolation of the electron 
identification efficiencies to energies above the Michel endpoint of 52.8 MeV. 
A GEANT 3.15-based Monte Carlo calculation is used for this purpose, as 
described in Section II. 
The MC generated events were checked against Michel data taken during the 
1994-1995 run periods. 
The electron ID efficiencies are determined in the MC Michel electron sample 
and in the data Michel electron sample. 
The differences observed in these two samples result in a 15\% uncertainty 
in the selection efficiency. 
When the Michel MC sample is compared to the DIF-MC sample a lower difference 
of 12\% is expected due to slightly narrower distributions in the DIF-MC sample.

The effect of the systematic uncertainties on the oscillation search can be 
explained as follows. 
The DIF oscillation search looks for an excess signal in the 
$\nu_e \, C \to e^- \, X$ process above the background from the $\nu_e$ 
contamination in the beam. 
This background flux is produced by the same DIF beam that produces the 
$\nu_\mu$ beam. 
The parent particle is dominantly either the $\mu^+$ or the $\pi^+$. 
The expected average number of events is given by 
\begin{equation}
\mu_{total} = 
\varepsilon \, \sigma_{\nu_e C} \, 
(\Phi_{\nu_\mu}P_{\nu_\mu \to \nu_e} + \Phi_{\nu_e} ) + \mu_{BUB}
\end{equation}
where $\varepsilon$ is the event selection efficiency, $\Phi_{\nu_e/\nu_\mu}$ 
are the neutrino fluxes, $\sigma_{\nu_e C}$ is the neutrino cross section, and 
$\mu_{BUB}$ is the cosmic-ray background. 
The oscillation signal is proportional to the same product, 
$\varepsilon \, \sigma_{\nu_e C} \, \Phi_{\nu_\mu}$, as the neutrino 
background, since $\Phi_{\nu_e}$ is proportional to $\Phi_{\nu_\mu}$. 
The effect of {\it lowering} the product 
$\varepsilon \, \Phi \, \sigma_{\nu_e C}$ 
is to reduce the predicted beam-related background, i.e. the background from 
neutrino interactions from the $\nu_e$ contamination in the beam. 
This raises the observed oscillation signal. 
Only by {\it raising} the product 
$\varepsilon \, \Phi \, \sigma_{\nu_e C}$ is the oscillation signal decreased.

In order to calculate conservative confidence regions, a value of the 
$\varepsilon \, \Phi \, \sigma_{\nu_e C}$ 21\% above the calculated value 
is assumed because of systematic uncertainties. 
This gives a larger confidence region because fewer of the excess events are 
attributed to oscillations. 
Then a value of $\varepsilon \, \Phi \, \sigma_{\nu_e C}$ 45\% lower than the 
calculated value is assumed, 
consistent with the LSND $\nu_\mu \, C$ measurement. 
This has the effect of moving the lower contour to the right. 
The upper contour also moves somewhat to the right. 
The final confidence region is shown as the logical OR of the two extreme 
cases just described.
\subsection{Confidence regions}

In order to determine the significance of the observed signal in terms 
of potential neutrino oscillation effects, a confidence level calculation 
is made in the context of a two-generation neutrino mixing model, as discussed 
in the Introduction. 
The oscillation probability is a function of the neutrino energy and the 
distance to the neutrino source. 
In the present case the distance to the source is ambiguous because of the
presence of multiple beam targets, A1, A2, and A6. 
Therefore, the energy distribution alone is used to determine the confidence 
levels in the $(\sin^2 2\theta,\Delta m^2)$ parameter space.  

The data are binned into four equal energy bins between 60 MeV and 200 MeV. 
In each bin the DIF-MC data are used to calculate the expected number of 
oscillation events, $\mu_{osc}$, and beam-related background events, 
$\mu_{BRB}$, at each $(\sin^2 2\theta,\Delta m^2)$ point. 
This number is added to the expected beam-unrelated background, $\mu_{BUB}$, 
to determine the total expected number of events in each of the four 
energy bins:
\begin{equation}
\vec{\mu}(\sin^2 2\theta,\Delta m^2) = 
\vec{\mu}_{osc}(\sin^2 2\theta,\Delta m^2)+\vec{\mu}_{BRB}+
                                           \vec{\mu}_{BUB}.
\end{equation}
From the expected numbers, a four-dimensional Poisson probability density, 
$p(\vec{N};\vec{\mu})$ (one dimension for each bin) of all possible results 
for this experiment is determined. 
An integration over all points in this space with a probability density 
greater than or equal to the measured data point value, 
\begin{equation}
{\cal P}(\vec{N};\vec{\mu}) \ \geq \ {\cal P}(\vec{N}_{meas};\vec{\mu}),
\end{equation}
gives a probability for the $(sin^2 2\theta,\Delta m^2)$ point:
\begin{equation}
\sum_{
p(\vec{N};\vec{\mu})\geq p(\vec{N}_{meas};\vec{\mu}) 
} p(\vec{N};\vec{\mu}) \, \, {\rm where} \, \vec{N}=(i,j,k,l), 
                                   \,  \, i,j,k,l=0,\ldots,\infty.
\end{equation}
This calculation determines confidence regions, or contours of equal 
probability, in the $(\sin^2 2\theta,\Delta m^2)$ space. 
As discussed above in Subsection A, the calculation is made for 
two extreme cases of the product $\varepsilon \, \sigma_{\nu_e C} \, \Phi$. 
The contours that result from the logical OR of these extremes are shown in 
Figure \ref{fig:confidence}. 
The calculation shows that the DIF result of this paper is consistent with 
the previous LSND DAR result \cite{bigpaper2} in terms of the two-generation 
oscillation parameters.
%
%
%
\section{Conclusions}
This paper reports a search for $\nu_e \, C \to e^- \, X$ interactions for 
electron energies $60 < E_e < 200$ MeV. 
Table \ref{tab:backgrounds} lists the expected contributions from conventional 
sources.
This search is motivated by a high sensitivity to neutrino oscillations of the 
type $\nu_\mu \to \nu_e$, due to the small contribution from conventional 
processes to the $\nu_e$ flux in this energy regime.
Two independent analyses observe a number of beam-on events significantly 
above the expected number from the sum of conventional beam-related processes 
and cosmic-ray (beam-off) events. 
The probability that the 12.5 (14.9) estimated background events fluctuate into 
23 (25) observed events is $7.0\times 10^{-3}$ ($1.6\times 10^{-2}$). 
The excess events are consistent with $\nu_\mu \to \nu_e$ oscillations
with an oscillation probability of $(2.6 \pm 1.0 \pm 0.5) \times 10^{-3}$. 
A fit to the event distributions, assuming neutrino oscillations as the source
of $\nu_e$, yields the allowed region in the $(\sin^2 2\theta,\Delta m^2)$ 
parameter space shown in Figure \ref{fig:confidence}. 
This allowed region is consistent with the allowed region from the DAR search 
reported earlier. 
This $\nu_\mu \to \nu_e$ DIF oscillation search has completely different 
backgrounds and systematic errors from the $\bar\nu_\mu \to \bar\nu_e$ DAR 
oscillation search and provides additional evidence that both effects are due 
to neutrino oscillations.
\paragraph*{Acknowledgments}
The authors gratefully acknowledge the support of Peter Barnes, 
Cyrus Hoffman, and John McClelland. 
This work was conducted under the auspices of the US Department of Energy, 
supported in part by funds provided by the University of California for 
the conduct of discretionary research by Los Alamos National Laboratory. 
This work is also supported by the National Science Foundation. 
We are particularly grateful for the extra effort that was made by these 
organizations to provide funds for running the accelerator at the end of 
the data taking period in 1995.
%
%
\clearpage
%
\begin{table}
\vspace*{\fill}
\caption{
Selection criteria for analysis A. 
For each criterion is listed: 
the value of the criterion; 
the efficiency of the criterion after all other criteria have been applied;
the number of events rejected by that criterion after all other criteria have 
been applied; 
the number of beam-on, beam-off and beam-excess events prior to applying the 
criterion.
}
\vspace{10mm}
\begin{tabular} {llrrrrr}
Criterion       & Cut Value  & Efficiency & $N_{rej}$ & On & Off &
         Excess \\ \hline
$F_{Cer}$       & $>$ 0.7    &       0.68 &       643 & 60 & 720 &
  9.6 $\pm$ 7.7 \\
$\Delta$        & $>$ 175 cm &       0.56 &       516 & 53 & 600 &
 11.4 $\pm$ 7.3 \\
$LR_{event}$    & $<$ 0.5    &       0.89 &       118 & 32 & 223 &
 16.7 $\pm$ 5.7 \\
Veto time       & $>$ 50 ns  &       0.98 &        27 & 26 & 138 &
 16.4 $\pm$ 5.1 \\
$cos\theta_\nu$ & $<$ 0.8    &       0.95 &        22 & 24 & 135 &
 14.7 $\pm$ 4.9 \\
Hot Spots       & (see text) &       0.94 &        22 & 25 & 134 &
 15.7 $\pm$ 4.9 \\
$L_{tc}$        & $<$ 1.1    &       0.98 &        11 & 23 & 125 &
 14.4 $\pm$ 4.8 \\
veto hits       & $<$ 3      &       0.98 &         3 & 23 & 117 &
 14.9 $\pm$ 4.8 \\
\end{tabular}
\label{tab:selection_a}
\end{table}
\newpage
%
\begin{table}
\vspace*{\fill}
\caption{
Selection criteria for analysis B. 
For each criterion is listed: 
the value of the criterion; 
the efficiency of the criterion after all other criteria have been applied;
the number of events rejected by that criterion after all other criteria have 
been applied; 
the number of beam-on, beam-off and beam-excess events prior to applying the 
criterion.
}
\vspace{10mm}
\begin{tabular} {llrrrrr}
Criterion       & Cut Value  & Efficiency & $N_{rej}$ & On &  Off &
         Excess \\ \hline
$S$             & $>$ 225 cm &       0.47 &      1009 & 89 & 1037 &
 16.8 $\pm$ 9.7 \\
$\rho$          & $>$ 0.4    &       0.83 &       689 & 68 &  738 &
 12.7 $\pm$ 8.5 \\
$L_{Tsci}$      & $<$ 1.15   &       0.90 &       131 & 34 &  214 &
 19.3 $\pm$ 5.9 \\
Veto time       & $>$ 70 ns  &       0.96 &        36 & 27 &  126 &
 18.0 $\pm$ 5.3 \\
$cos\theta_\nu$ & $<$ 0.8    &       0.95 &        14 & 27 &  104 &
 19.7 $\pm$ 5.2 \\
$L_{Tcer}$      & $<$ 1.5    &       0.98 &        10 & 26 &  101 &
 18.9 $\pm$ 4.1 \\
\end{tabular}
\label{tab:selection_b}
\end{table}
\newpage
%
\begin{table}
\vspace*{\fill}
\caption{
Event count after all selection criteria have been applied (analyses A/B). 
$Q_{A6}$ is the number of protons on target in Coulombs.
}
\vspace{10mm}
\begin{tabular} {cccccc}
 Year & $Q_{A6}$ & Beam On (A/B) & Beam Off (A/B) & Duty ratio &
 Excess (A/B)                    \\ \hline
 1993 & 1787     &   1 /  2      &   17 / 21      & 0.072      &
 -0.2 $\pm$ 1.0 /  0.5 $\pm$ 1.5 \\
 1994 & 5904     &  12 / 15      &   42 / 41      & 0.078      &
  8.7 $\pm$ 3.5 / 11.8 $\pm$ 3.9 \\
 1995 & 7081     &  10 /  8      &   55 / 30      & 0.060      &
  6.7 $\pm$ 3.2 /  6.2 $\pm$ 2.8 \\
Total & 14772    &  23 / 25      &  114 / 92      & 0.070      &
 15.1 $\pm$ 4.9 / 18.5 $\pm$ 5.0 \\
\end{tabular}
\label{tab:events} 
\end{table}
%
\vspace{20mm}
\begin{table}
\caption{
Backgrounds and observed numbers of beam-on events. 
The beam-unrelated cosmic backgrounds are calculated in two ways. 
The product of the beam duty ratio and the number of beam-off events is shown 
first.
The correponding result for the fitted number of beam-unrelated background 
events
in the beam-on sample is shown in parenthesis. 
The distinction is described in the text.
}
\vspace{10mm}
\begin{tabular} {ccc}
                                 & Analysis A                              &
   Analysis B                           \\ \hline
Beam-unrelated background        &   8.0 $\pm$ 0.7 (6.2 $\pm$ 2.0)         &
  6.4 $\pm$ 0.7  (5.9 $\pm$ 1.9)        \\
Expected beam-related background &   4.5 $\pm$ 0.9                         &
  8.5 $\pm$ 1.7                         \\ 
Total expected background        &  12.5 $\pm$ 1.1 (10.7 $\pm$ 2.2)        &
 14.9 $\pm$ 1.8  (14.4 $\pm$ 2.6)       \\
Observed beam-on events          &  23                                     &
 25                                     \\
Fluctuation probability          &  $7.0\times10^{-3}$($1.2\times10^{-3}$) &
 $1.6\times10^{-2}$ ($1.2\times10^{-2}$)\\
\end{tabular}
\label{tab:fitback}
\end{table}
\newpage
%
\begin{table}
\vspace*{\fill}
\caption{
Comparison of results for the A, B, AND, and OR data sets. 
The BUB is the beam-unrelated background from cosmic rays.
}
\vspace{10mm}
\begin{tabular}{ccccccc}
Sample     & Beam On/Off & BUB            & $\nu$-Background & Osc. Excess    &
 Efficiency & Osc. Probability  \\ \hline
Analysis A &    23/114   & $ 8.0 \pm 0.7$ & $4.5 \pm 0.9$    & $10.5 \pm 4.9$ &
  0.084     & $(2.9 \pm 1.4)\times 10^{-3}$ \\
Analysis B &    25/ 92   & $ 6.4 \pm 0.7$ & $8.5 \pm 1.7$    & $10.1 \pm 5.3$ &
  0.138     & $(1.7 \pm 0.9)\times 10^{-3}$ \\
AND        &     8/ 31   & $ 2.2 \pm 0.3$ & $3.1 \pm 0.6$    & $ 2.7 \pm 2.9$ &
  0.055     & $(1.1 \pm 1.2)\times 10^{-3}$ \\
OR         &    40/175   & $12.3 \pm 0.9$ & $9.6 \pm 1.9$    & $18.1 \pm 6.6$ &
  0.165     & $(2.6 \pm 1.0)\times 10^{-3}$ \\
\end{tabular}
\label{tab:andorisgm}
\end{table}
%
\vspace{20mm}
\begin{table}
\caption{
The background estimates for the $\nu_\mu \to \nu_e$ oscillation search are 
shown
for a $d > 0$ detector fiducial volume, $9.2 \times 10^{22}$ protons on target, 
and for reconstructed energies between 60 MeV and 200 MeV. 
The number of events for 100\% $\nu_\mu \to \nu_e$ transmutation is shown also. 
These numbers are illustrative for an electron selection efficiency of 0.10, 
independent of energy. 
The actual efficiencies in analyses A and B are slightly different and energy 
dependent.
}
\vspace{10mm}
\begin{tabular}{lcccc}
Process & Flux (cm$^{-2}/$POT)          & $<\sigma>_\nu$ ($10^{-40}$ cm$^2$) &
 Eff.    & Number of Events \\ \hline
$\nu_e   C \to e^- X$ ($\mu$ DIF)       & $3.8 \times 10^{-14}$              &
 28.3    & 0.10  &  $3.8$   \\ 
$\nu_e   C \to e^- X$ ($\pi$ DIF)       & $8.3 \times 10^{-15}$              &
 79.2    & 0.10  &  $1.6$   \\
$\nu_\mu C \to \nu_\mu C \pi^{\circ}$   & $6.5 \times 10^{-11}$              &
 1.6     & 0.06  &  $0.3$   \\
$\nu_\mu e \to \nu_\mu e$               & $6.5 \times 10^{-11}$              &
 0.00136 & 0.005 &  $0.1$   \\ \hline
Total background                        &                                    &
         &       &  $5.8$   \\ \hline
100\% $\nu_\mu \to \nu_e$ transmutation &                                    &
         &       &  $4470$  \\
\end{tabular}
\label{tab:backgrounds}
\end{table}
\clearpage
%
\begin{figure}[p]
\vspace*{\fill}
\begin{center}
\mbox{
\epsfxsize=16.0cm \epsfbox[ 1   1 430 430]{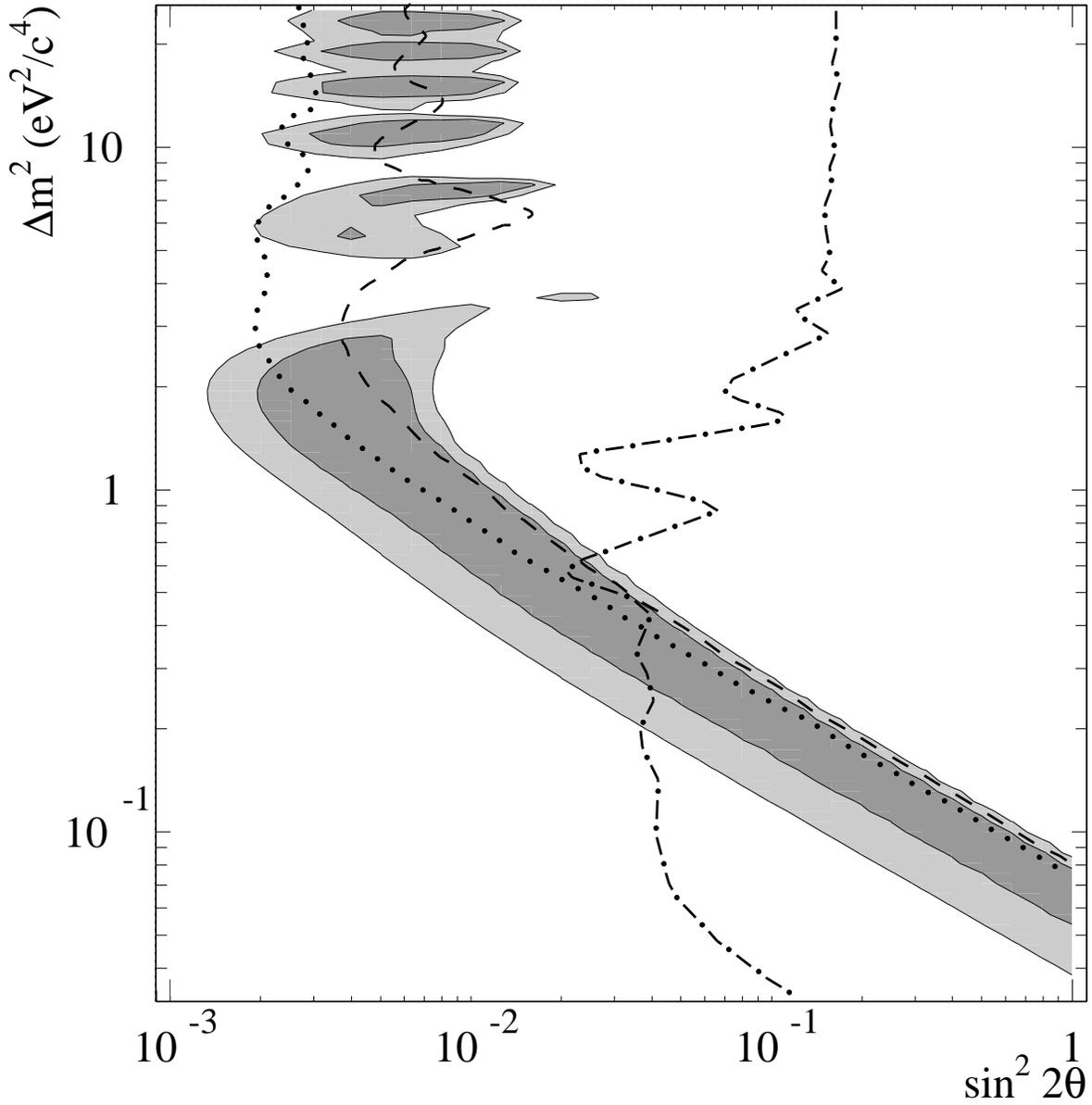}
}
\end{center}
\vspace{5mm}
\caption{
The LSND $(\sin^2 2\theta,\Delta m^2)$ favored regions obtained from the 
$\bar\nu_\mu \to \bar\nu_e$ DAR oscillations search. 
The darkly-shaded and lightly-shaded regions correspond to 90\% and 99\% 
likelihood regions. 
Also shown are the 90\% confidence level limits from KARMEN (dashed), E776 
(dotted) and the Bugey reactor experiment (dot-dashed).
}
\label{fig:loglik}
\end{figure}
\newpage
%
\begin{figure}[p]
\vspace*{\fill}
\begin{center}
\mbox{
\epsfxsize=16.0cm \epsfbox[20  20 550 550]{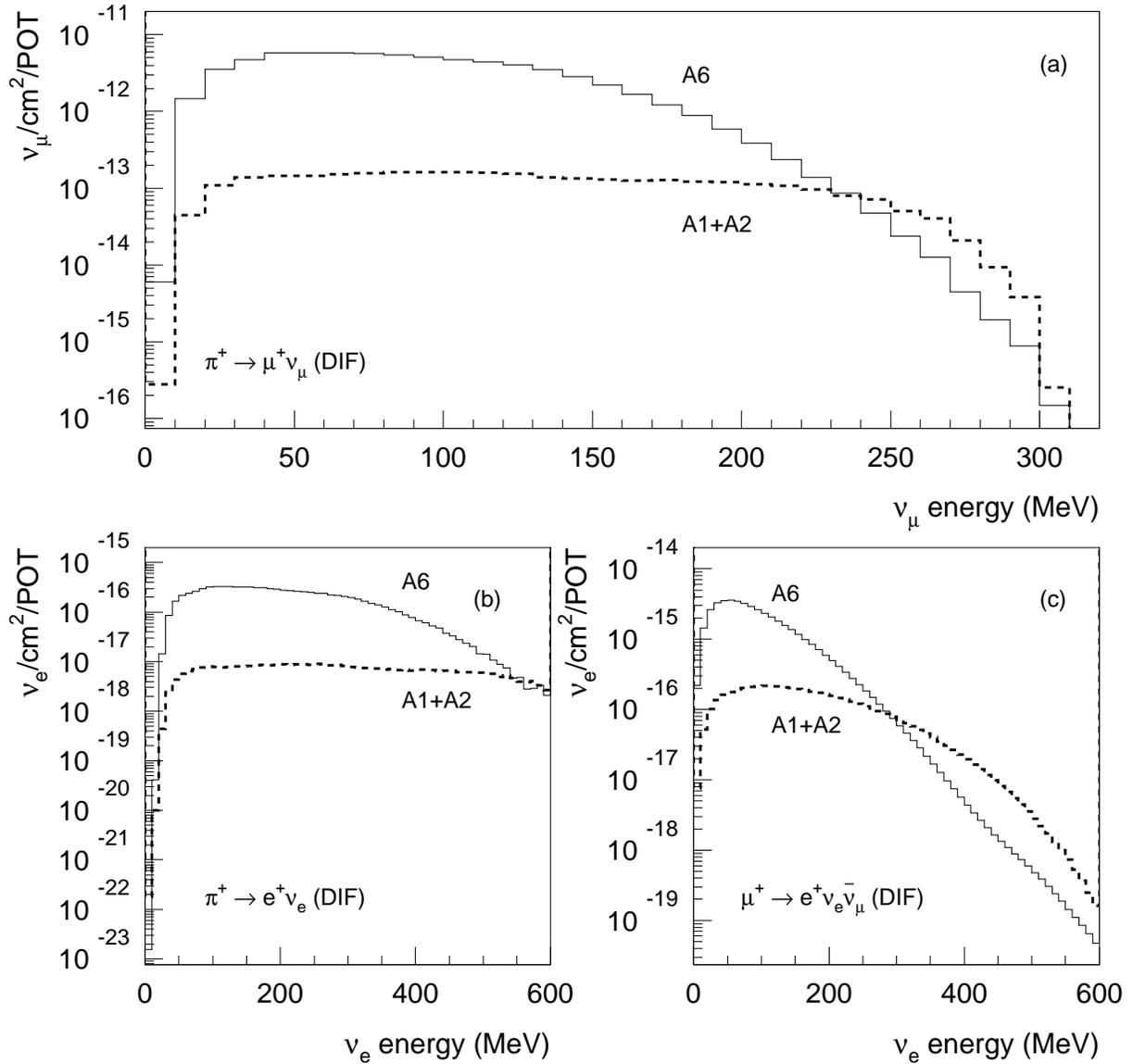}
}
\end{center}
\vspace{5mm}
\caption{
Calculated $\nu_\mu$ and $\nu_e$ DIF fluxes at the detector center from the A6 
target (solid histograms) and from the A1+A2 targets (dashed histograms).
}
\label{fig:a126_flux94}
\end{figure}
\newpage
\begin{figure}[p]
\begin{center}
\mbox{
\epsfxsize=16.0cm \epsfbox[20 270 550 550]{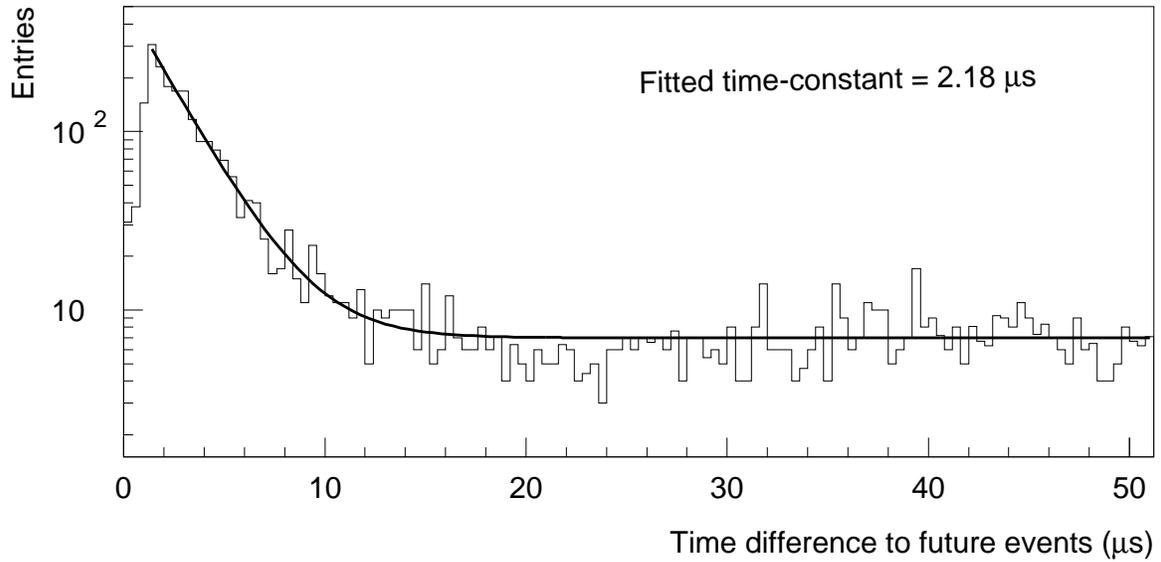}
}
\end{center}
\vspace{1mm}
\caption{
Time difference distribution to events subsequent to all primary events for 
the initial DIF data after standard PID, veto shield hit multiplicity and 
fiducial volume cuts.
}
\label{fig:fdt9495}
\end{figure}
%
\begin{figure}[p]
\begin{center}
\mbox{
\epsfxsize=16.0cm \epsfbox[20 270 550 550]{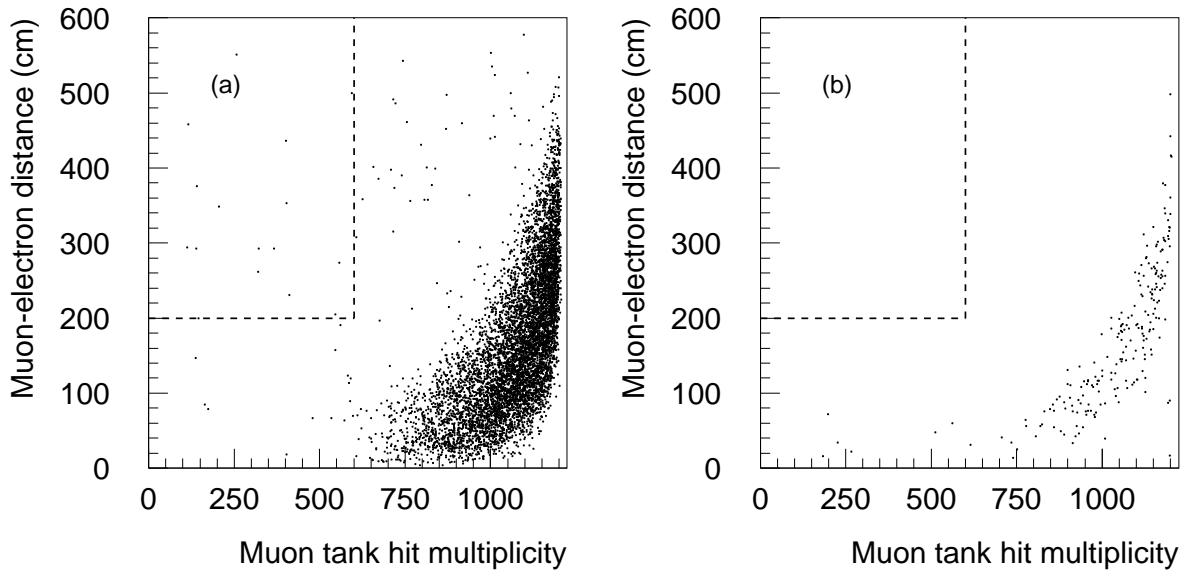}
}
\end{center}
\vspace{1mm}
\caption{
Reconstructed distance distribution between Michel electron and parent 
stopped cosmic-ray muon, versus muon tank hit multiplicity. 
Figure (a) is for muons with a veto shield hit multiplicity $\ge$ 6 and 
(b) is for muons with a veto shield hit multiplicity $<$ 6. 
The regions in the upper left corners, delimited by the dashed lines, are 
the regions allowed by this selection (see text).
}
\label{fig:prmuon}
\end{figure}
\newpage
%
\begin{figure}[p]
\vspace*{\fill}
\begin{center}
\mbox{
\epsfxsize=16.0cm \epsfbox[20 270 550 550]{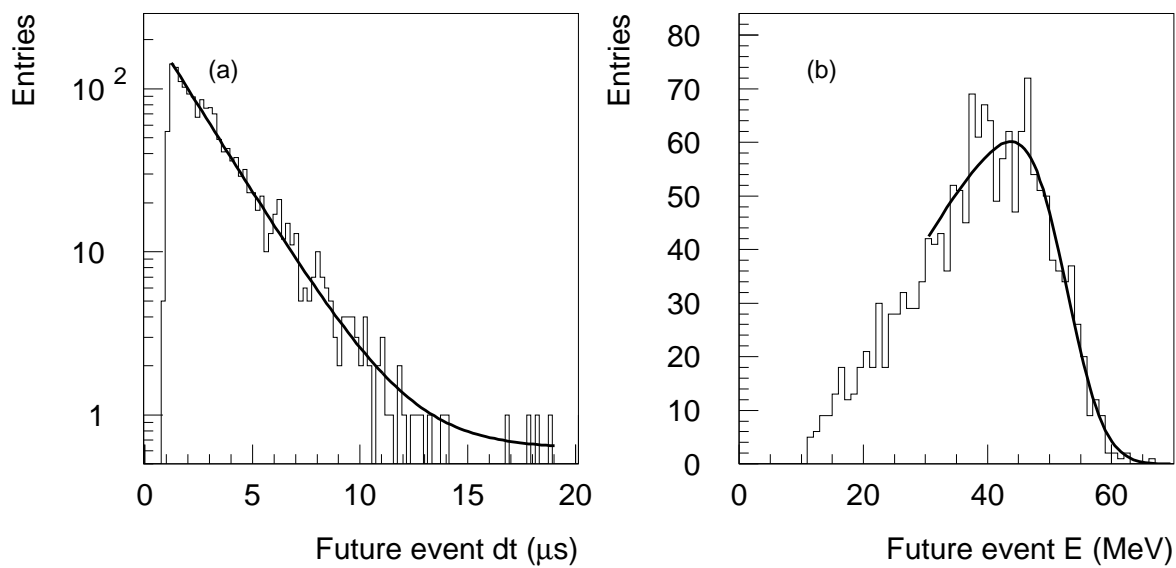}
}
\end{center}
\vspace{5mm}
\caption{
(a) Time difference to future events and (b) energy distribution of the future 
events for primary events of the initial DIF sample that fail the future 
space-time correlations. 
The fit in (a) is to an exponential plus a constant with a time constant of 
2.2 $\mu s$. 
The fit in (b) is to the Michel electron spectrum shape.
}
\label{fig:future_te}
\end{figure}
\newpage
%
\begin{figure}[p]
\vspace*{\fill}
\begin{center}
\mbox{
\epsfxsize=16.0cm \epsfbox[20  20 550 550]{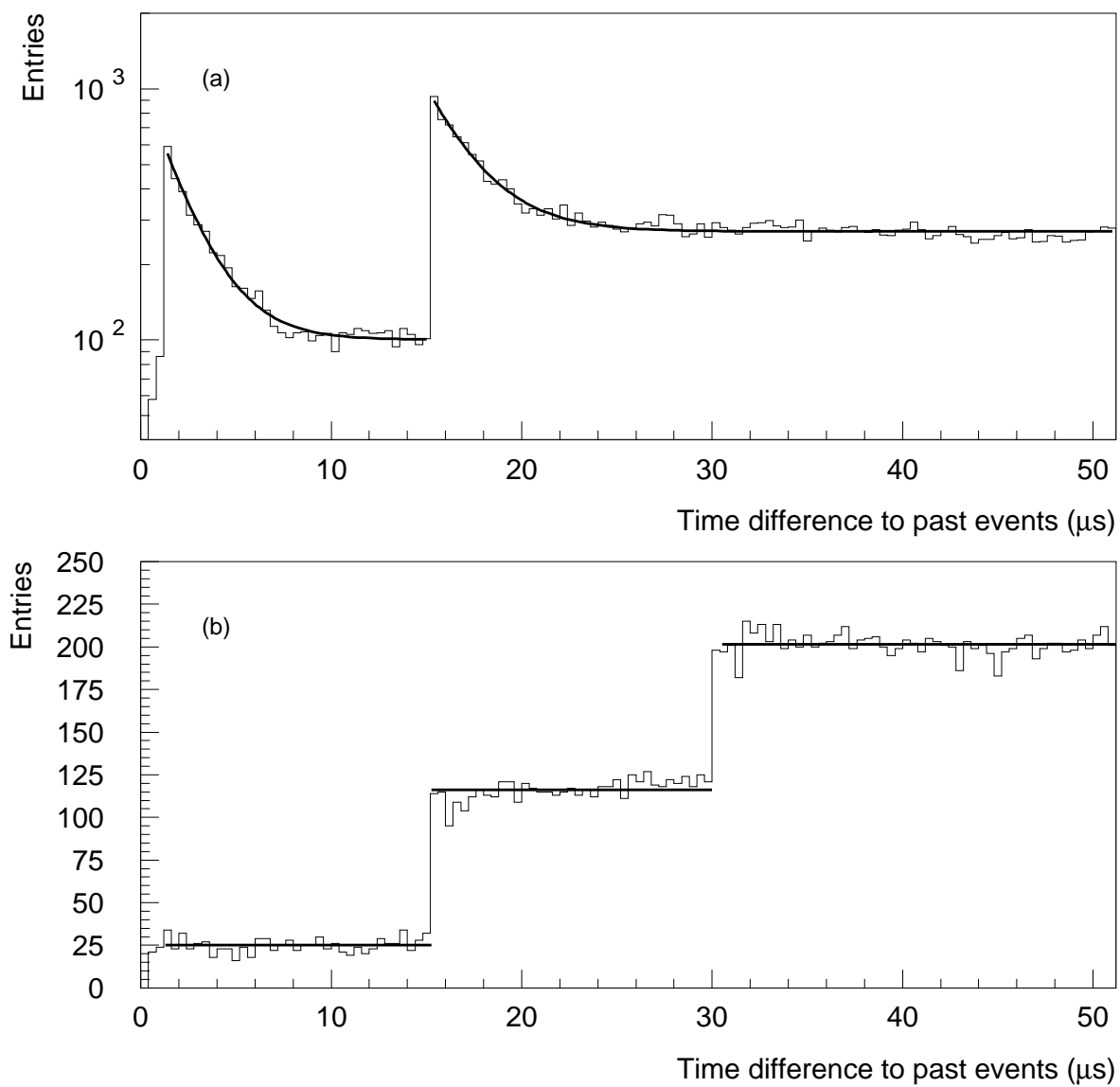}
}
\end{center}
\vspace{5mm}
\caption{
Time difference to all previous events (a) before and (b) after the past 
space-time correlation cuts. 
The sharp edge at 15.2 $\mu$s in both (a) and (b) is a reflection of the DAQ 
operation, as described in the text. 
The sharp edge at 30 $\mu$s in (b) is induced by the selection algorithm.
}
\label{fig:pastdt}
\end{figure}
\newpage
%
\begin{figure}[p]
\vspace*{\fill}
\begin{center}
\mbox{
\epsfxsize=16.0cm \epsfbox[20  20 550 550]{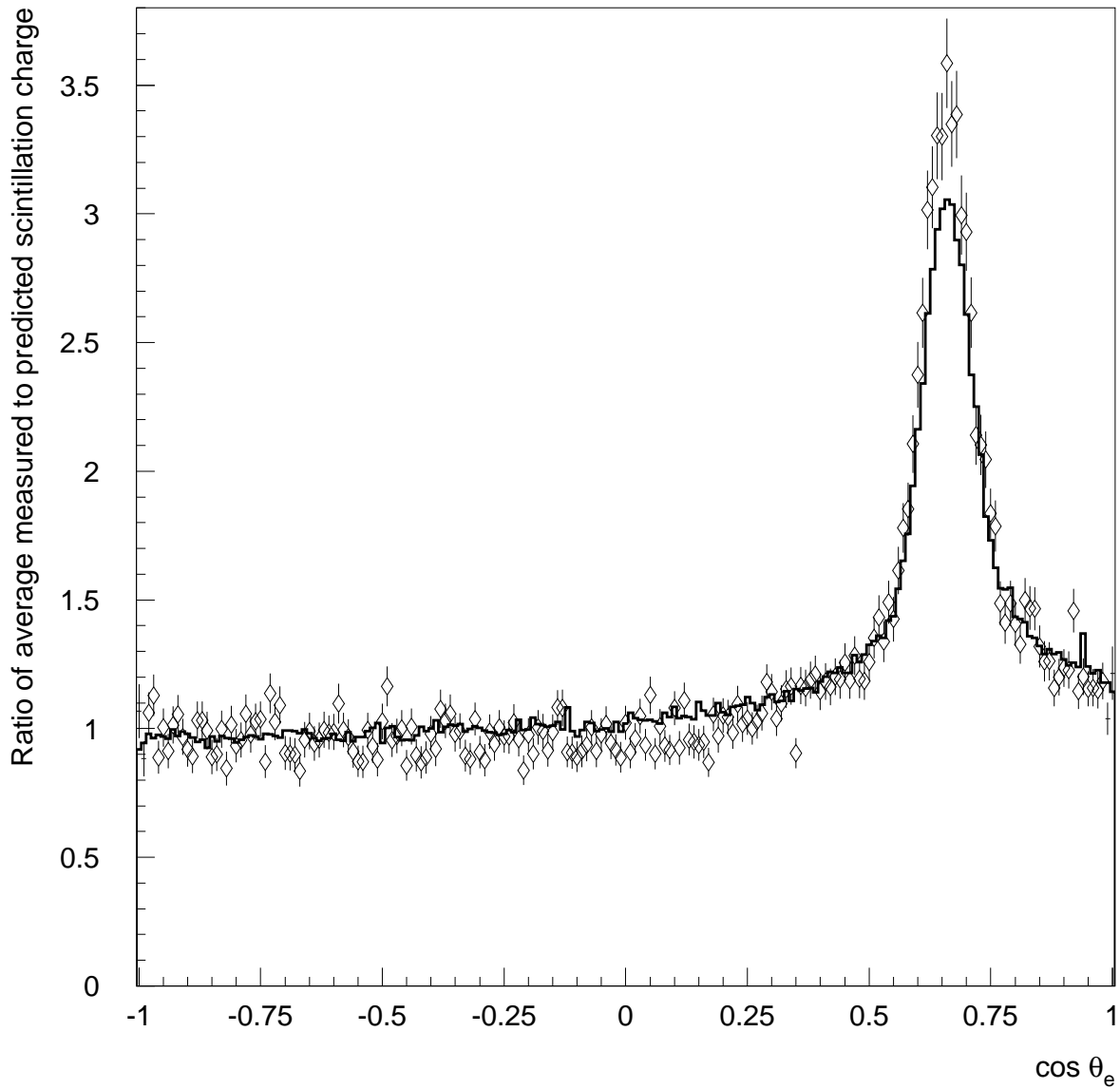}
}
\end{center}
\vspace{5mm}
\caption{
The ratio of the average measured charge to the predicted scintillation charge 
versus cos $\theta_e$, the cosine of the angle between the event direction and 
the PMTs. 
The solid histogram shows the data, and the points with error bars show the MC 
simulation.
}
\label{fig:cercosth_me_mc95}
\end{figure}
\newpage
%
\begin{figure}[p]
\vspace*{\fill}
\begin{center}
\mbox{
\epsfxsize=16.0cm \epsfbox[20  20 550 550]{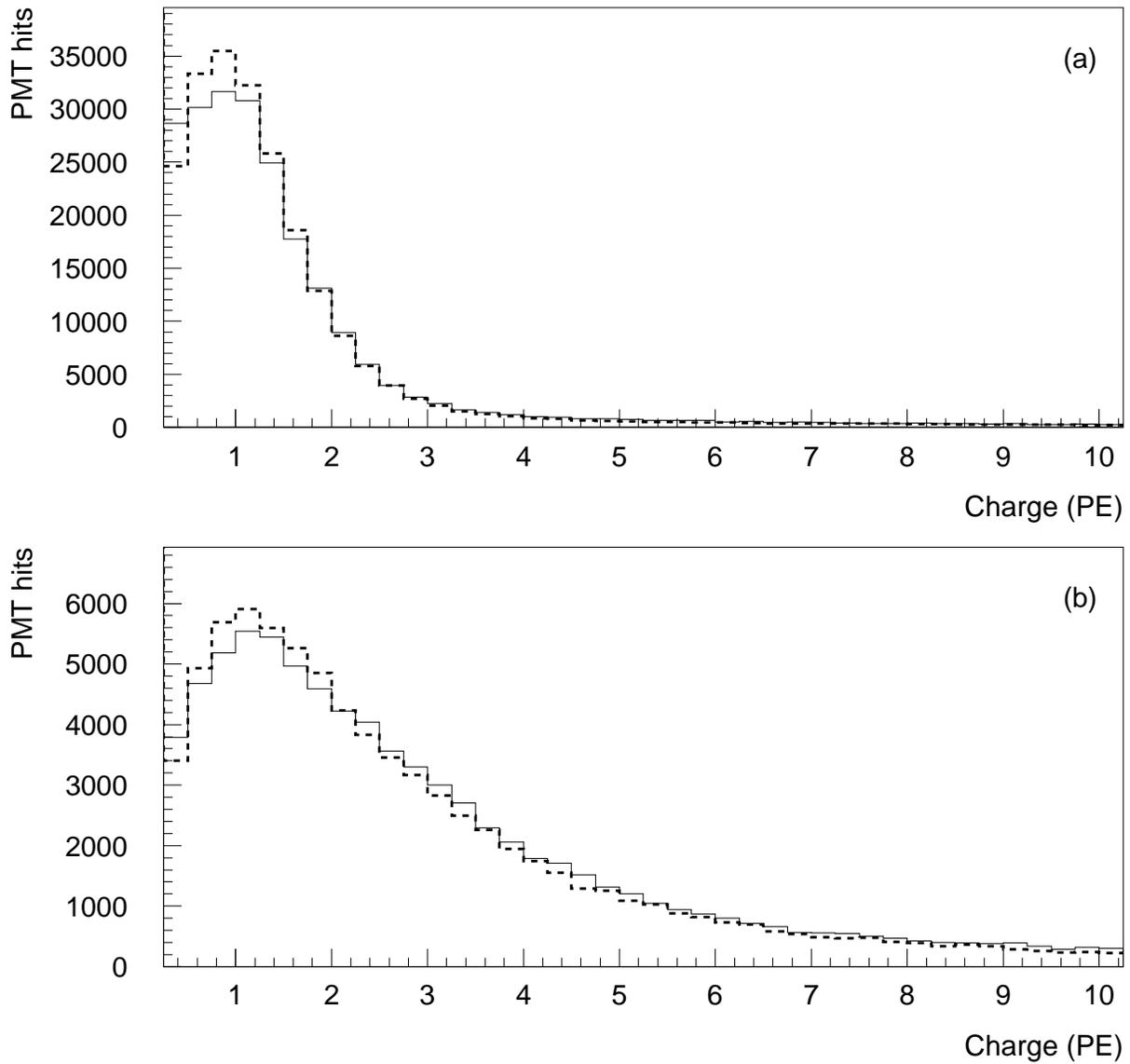}
}
\end{center}
\vspace{5mm}
\caption{
Unnormalized charge response functions of the PMTs for two values of the 
predicted charge (a) $\mu$ = 0.0-0.5 PE and (b) $\mu$ = 2.5-3.0 PE. 
The solid histograms show the data, and the dashed histograms show the MC 
simulation.
}
\label{fig:chargelike}
\end{figure}
\newpage
%
\begin{figure}[p]
\vspace*{\fill}
\begin{center}
\mbox{
\epsfxsize=16.0cm \epsfbox[20  20 550 550]{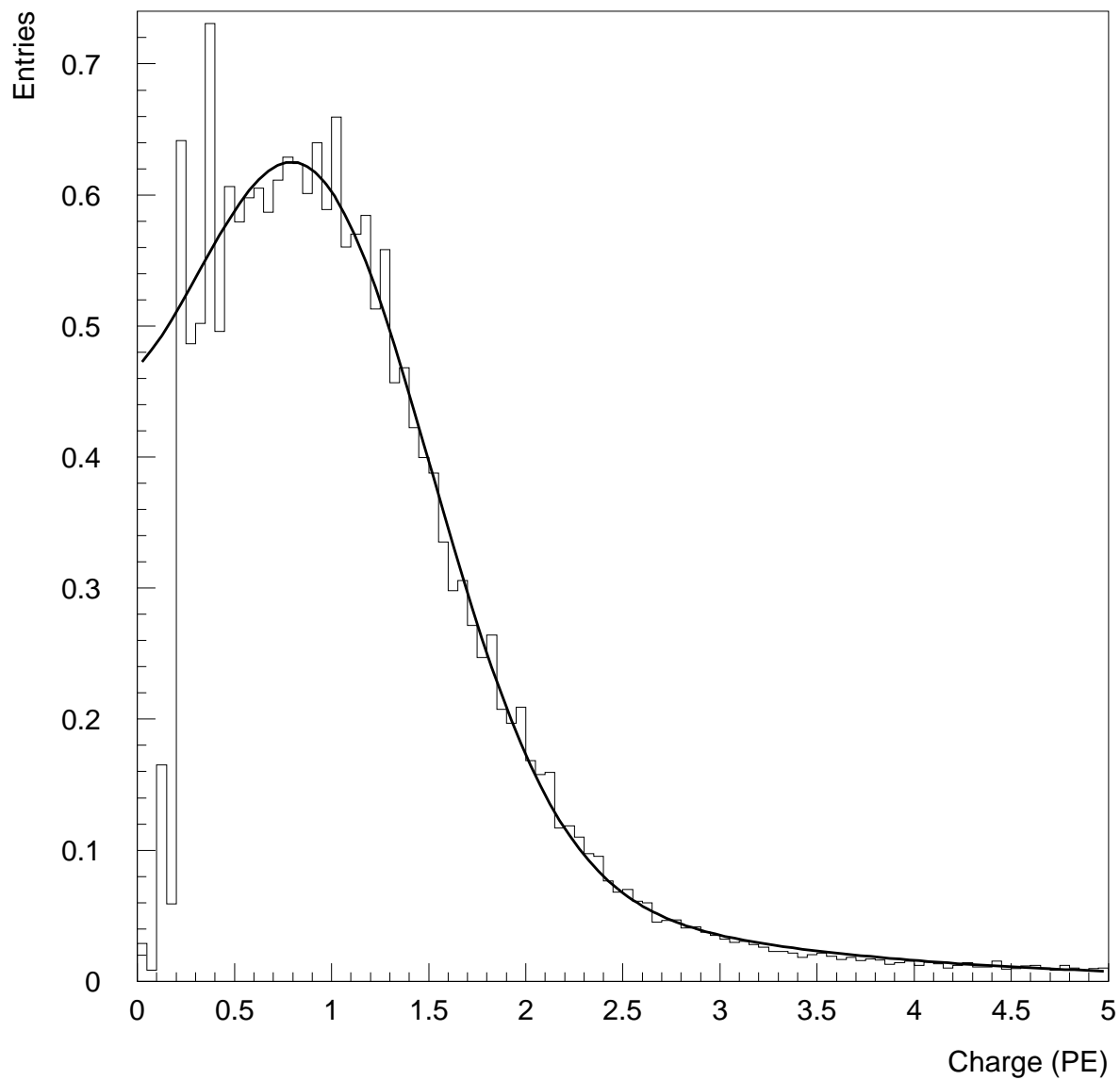}
}
\end{center}
\vspace{5mm}
\caption{
Single-PE charge response function of the PMTs. 
The distribution is fitted to a Gaussian plus an exponential and is normalized 
to unit area. 
The 0.2 PE threshold is clearly visible.
}
\label{fig:pq1_95}
\end{figure}
\newpage
%
\begin{figure}[p]
\vspace*{\fill}
\begin{center}
\mbox{
\epsfxsize=16.0cm \epsfbox[20  20 550 550]{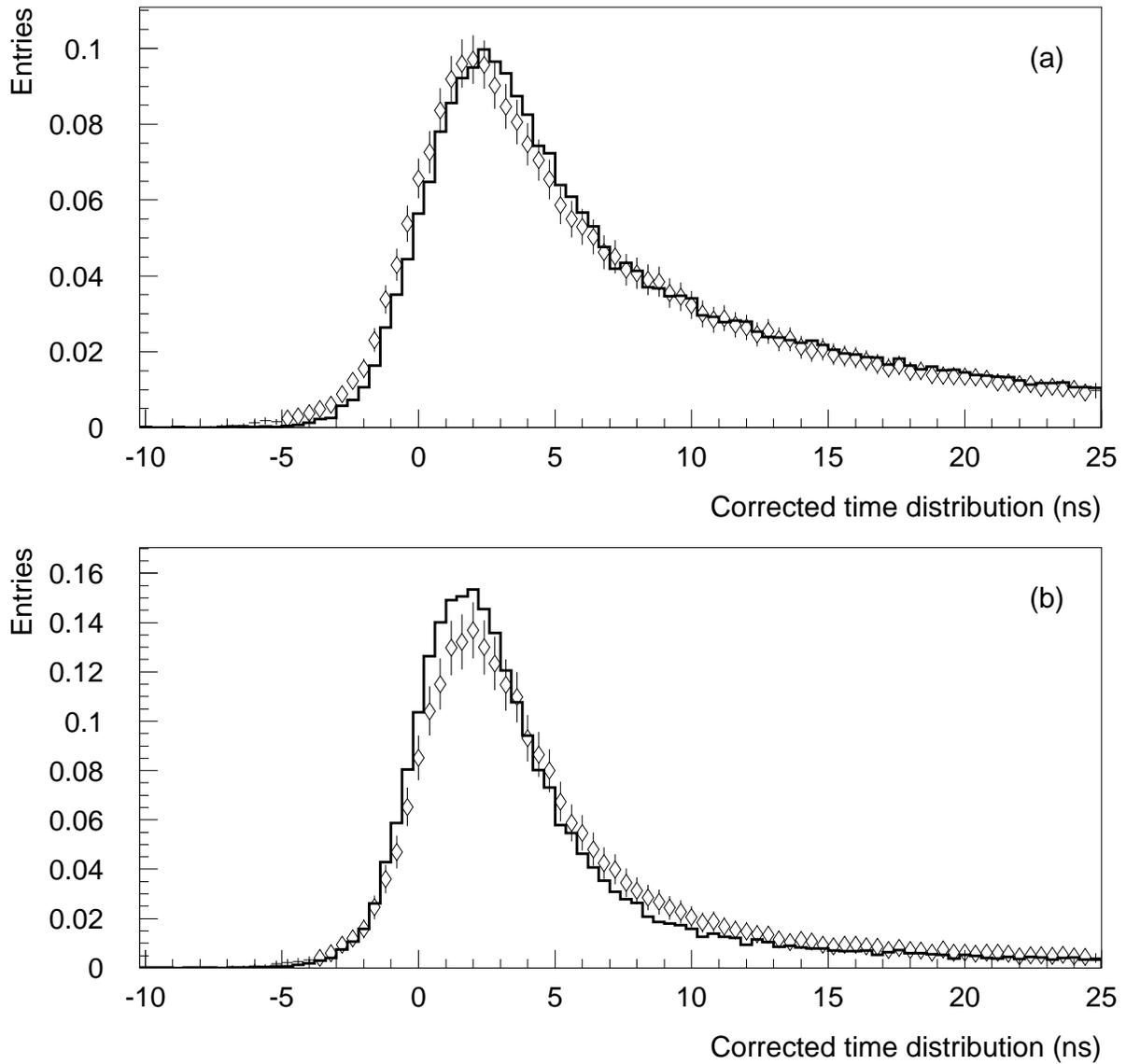}
}
\end{center}
\vspace{5mm}
\caption{
Time response functions of the PMTs for two values of the predicted 
charge (a) $\mu$ = 0.0-1.0 PE and (b) $\mu$ = 5.0-6.0 PE for the isotropic + 
scattered $\check {\rm C}$erenkov light components in Michel electron events. 
The solid histograms show the data, and the points with error bars show the MC 
simulation.
}
\label{fig:timelike_a}
\end{figure}
\newpage
%
\begin{figure}[p]
\vspace*{\fill}
\begin{center}
\mbox{
\epsfxsize=16.0cm \epsfbox[20  20 550 500]{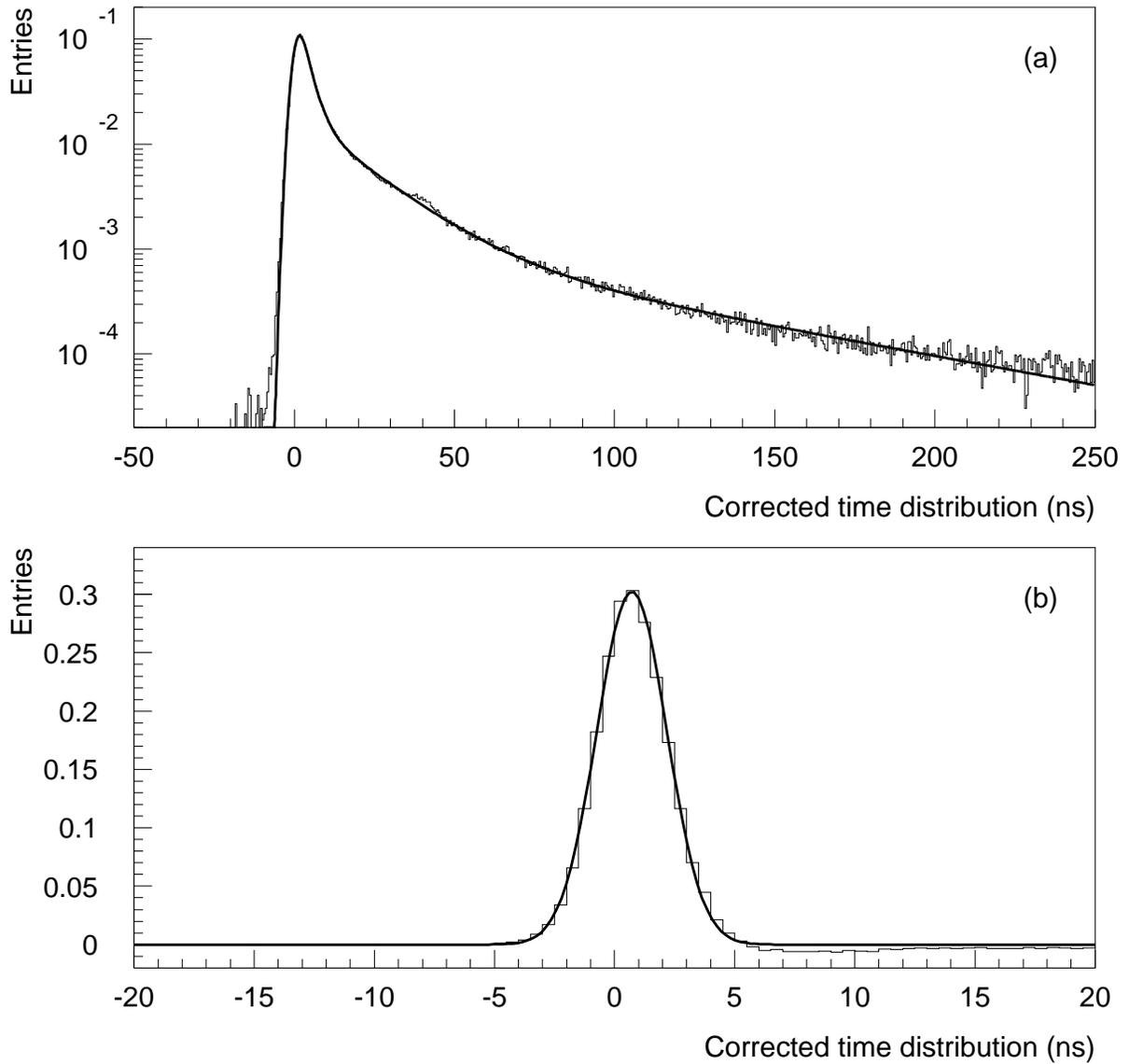}
}
\end{center}
\vspace{5mm}
\caption{
Corrected time distributions from ``monoenergetic'' data Michel electrons for 
(a) scintillation + scattered $\check {\rm C}$erenkov light and (b) direct 
$\check {\rm C}$erenkov light. 
The fit in (a) is to a convolution of a Gaussian with the sum of three 
exponentials and in (b) to a Gaussian. 
Both distributions are normalized to unit area.
}
\label{fig:tcbwdcer_95}
\end{figure}
\newpage
%
\begin{figure}[p]
\vspace*{\fill}
\begin{center}
\mbox{
\epsfxsize=16.0cm \epsfbox[20  20 550 500]{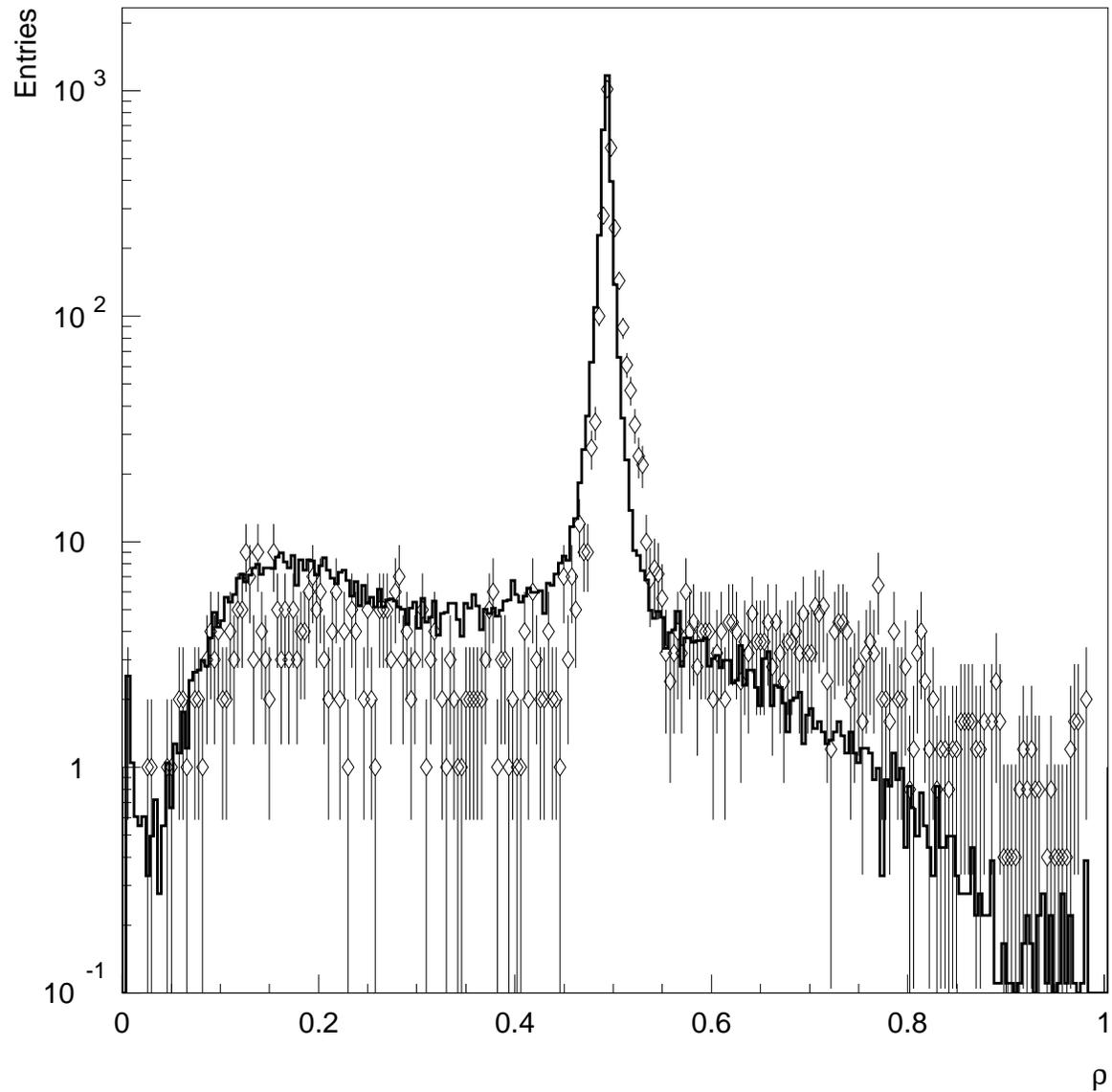}
}
\end{center}
\vspace{5mm}
\caption{
$\check {\rm C}$erenkov-to-scintillation density ratio $\rho$ for 
``monoenergetic'' Michel electron events. 
The solid histogram shows the data, and the points with error bars show the MC 
simulation.
}
\label{fig:fcsr_me_mc95}
\end{figure}
\newpage
%
\begin{figure}[p]
\vspace*{\fill}
\begin{center}
\mbox{
\epsfxsize=16.0cm \epsfbox[20 270 550 500]{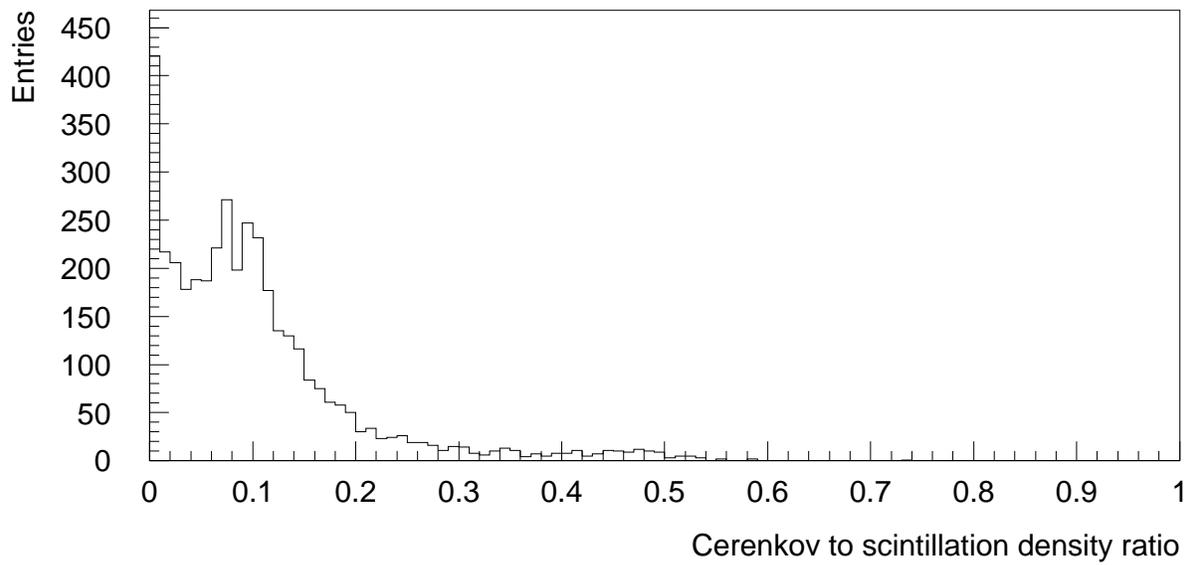}
}
\end{center}
\vspace{5mm}
\caption{
$\check{\rm C}$erenkov-to-scintillation density ratio $\rho$ for cosmic-ray 
neutron events with electron-equivalent energies between 60 and 200 MeV.
}
\label{fig:neutrons_fcsr}
\end{figure}
\newpage
%
\begin{figure}[p]
\vspace*{\fill}
\begin{center}
\mbox{
\epsfxsize=16.0cm \epsfbox[20  20 550 550]{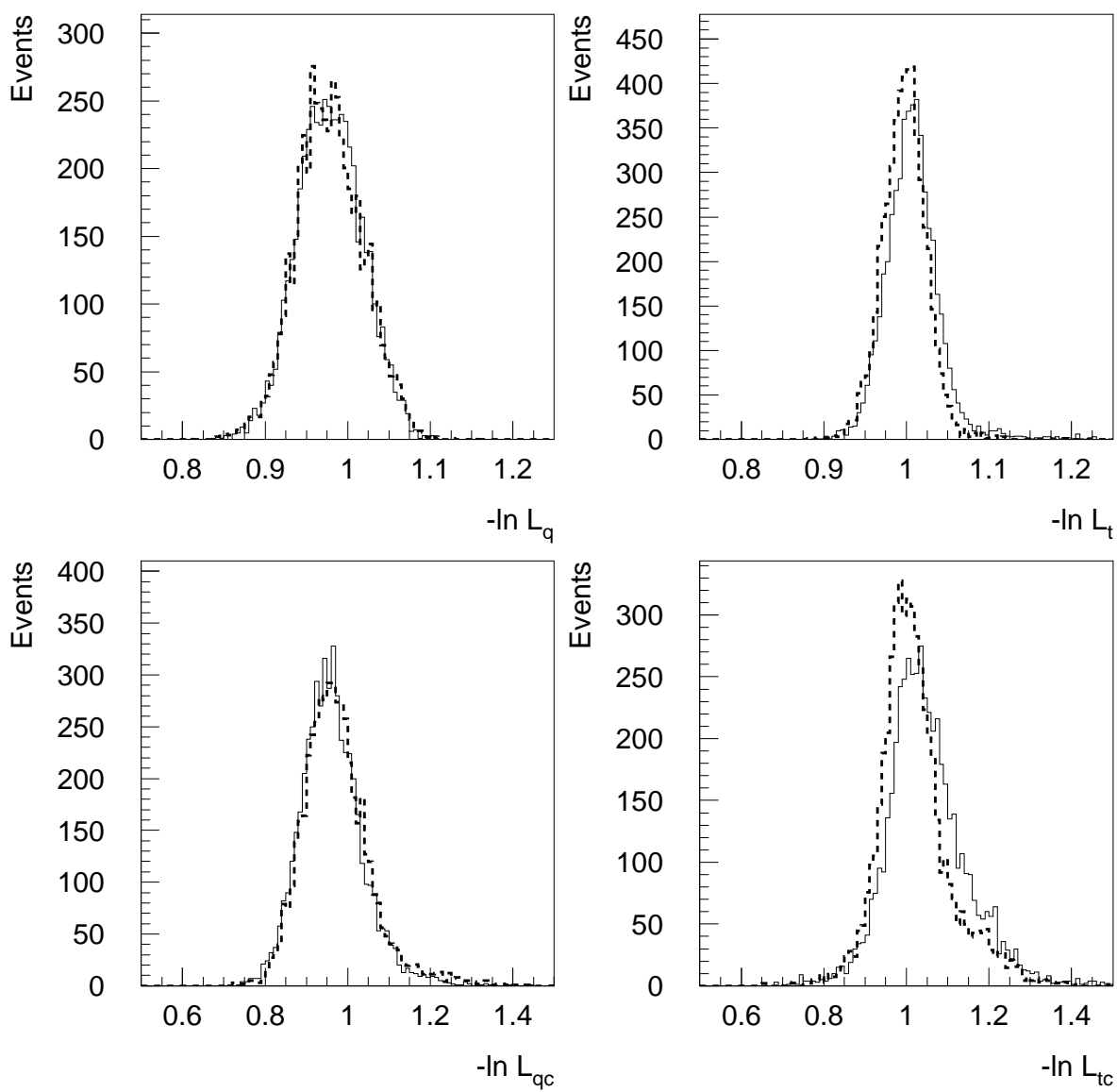}
}
\end{center}
\vspace{5mm}
\caption{
Charge and time negative log-likelihoods for Michel electrons, as calculated 
for the entire event (top) and in the $\check {\rm C}$erenkov cone only 
(bottom) - analysis A. 
The solid histograms show the data, and the dashed histograms show the MC 
simulation.
}
\label{fig:michellike}
\end{figure}
\newpage
%
\begin{figure}[p]
\vspace*{\fill}
\begin{center}
\mbox{
\epsfxsize=16.0cm \epsfbox[20  20 550 550]{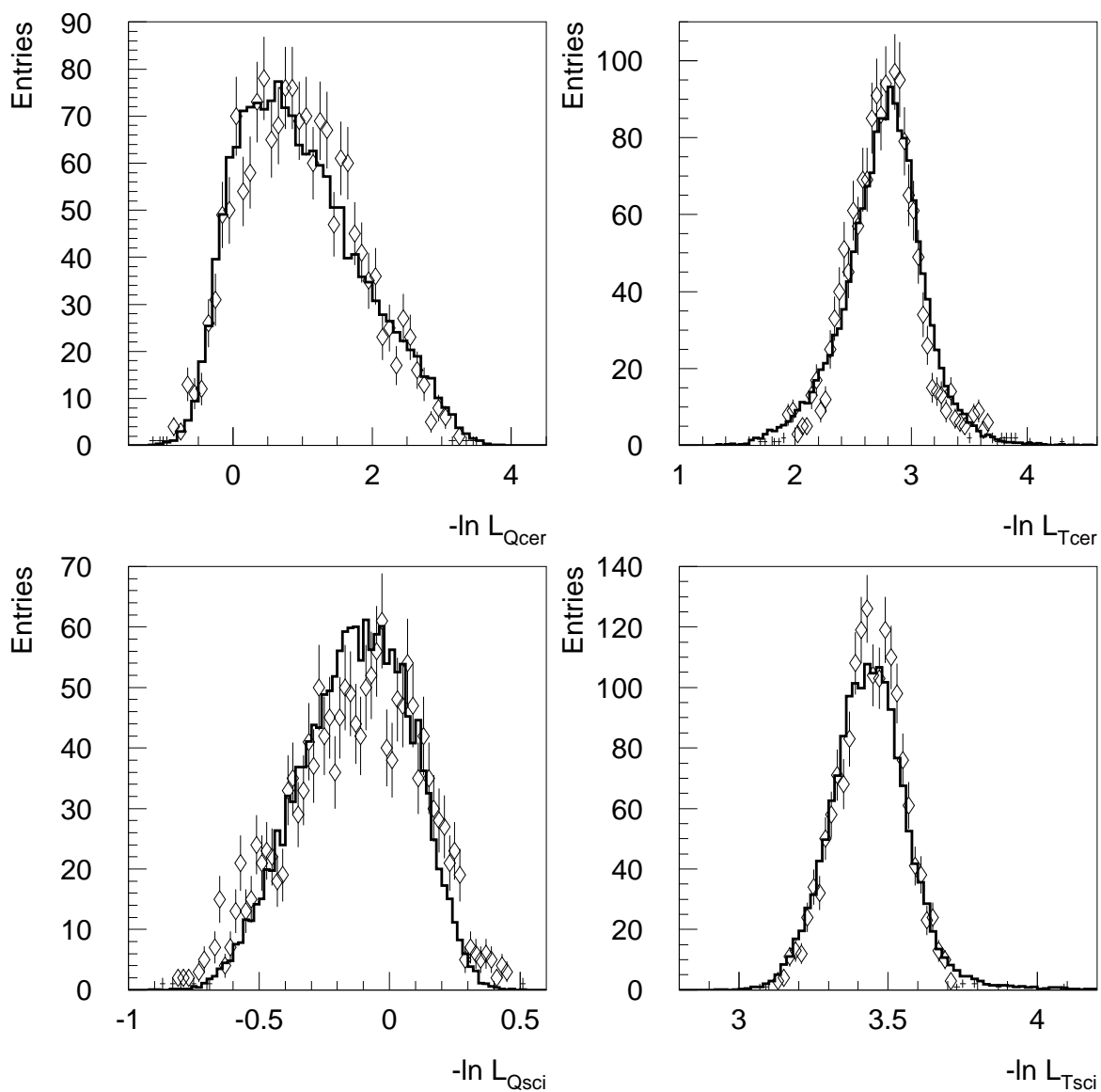}
}
\end{center}
\vspace{5mm}
\caption{
Charge and time negative log-likelihoods for ``monoenergetic'' Michel 
electrons, inside (top) and outside (bottom) the $\check {\rm C}$erenkov cone 
- analysis B. 
The solid histograms show the data, and the points with error bars show the MC 
simulation.
}
\label{fig:llq_me_mc95}
\end{figure}
\newpage
%
\begin{figure}[p]
\begin{center}
\mbox{
\epsfxsize=16.0cm \epsfbox[20 270 550 550]{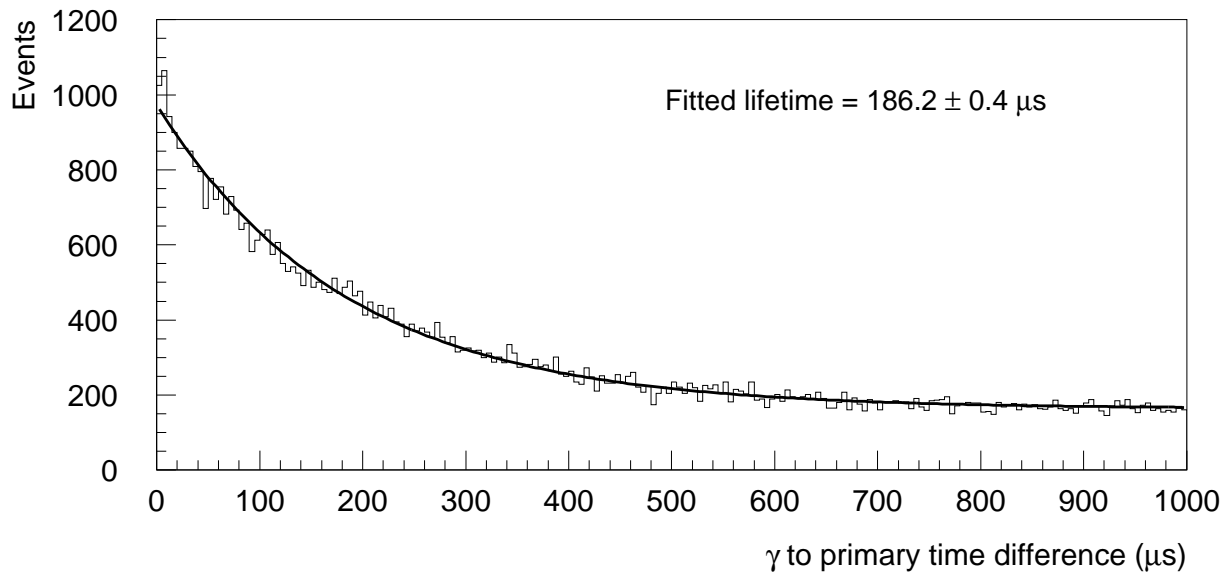}
}
\end{center}
\vspace{5mm}
\caption{
Time difference distribution between the photons and the primary events in 
the initial DIF (beam on + off) data sample. 
The fit is to an exponential plus a constant.
}
\label{fig:gdt_all9495}
\end{figure}
%
\begin{figure}[p]
\begin{center}
\mbox{
\epsfxsize=16.0cm \epsfbox[20 270 550 550]{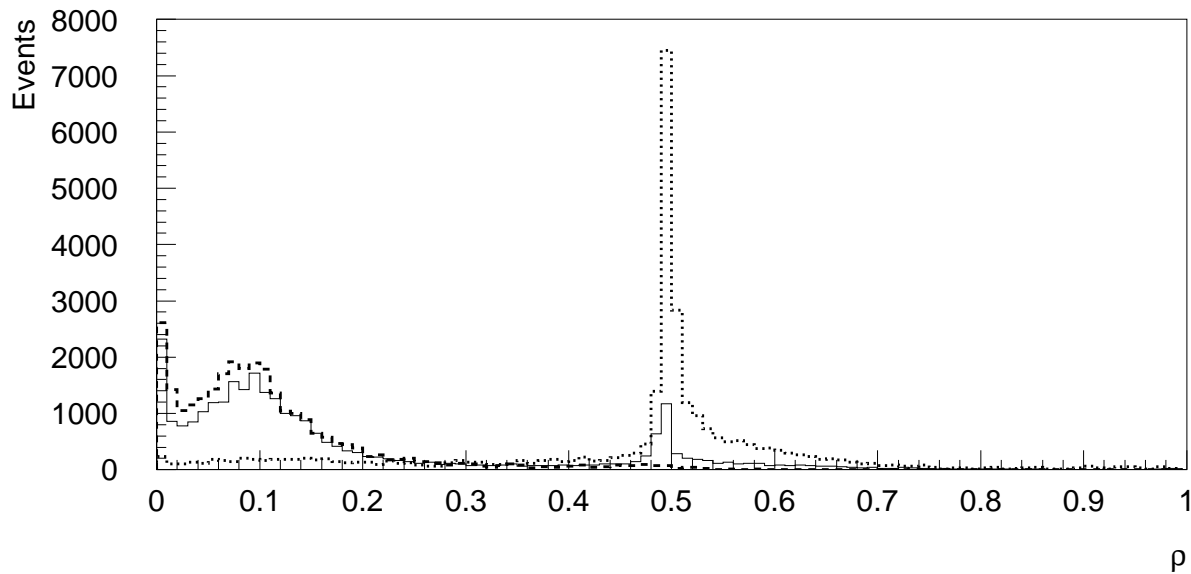}
}
\end{center}
\vspace{5mm}
\caption{
$\check {\rm C}$erenkov-to-scintillation density ratio $\rho$ for all DIF data 
(beam on+off) for analysis B. 
Superimposed are the same distributions for cosmic-ray neutrons (dashed) 
and for the DIF-MC electron sample (dotted), normalized to the same area.
}
\label{fig:fcsr_difall}
\end{figure}
\newpage
%
\begin{figure}[p]
\vspace*{\fill}
\begin{center}
\mbox{
\epsfxsize=16.0cm \epsfbox[20  20 550 550]{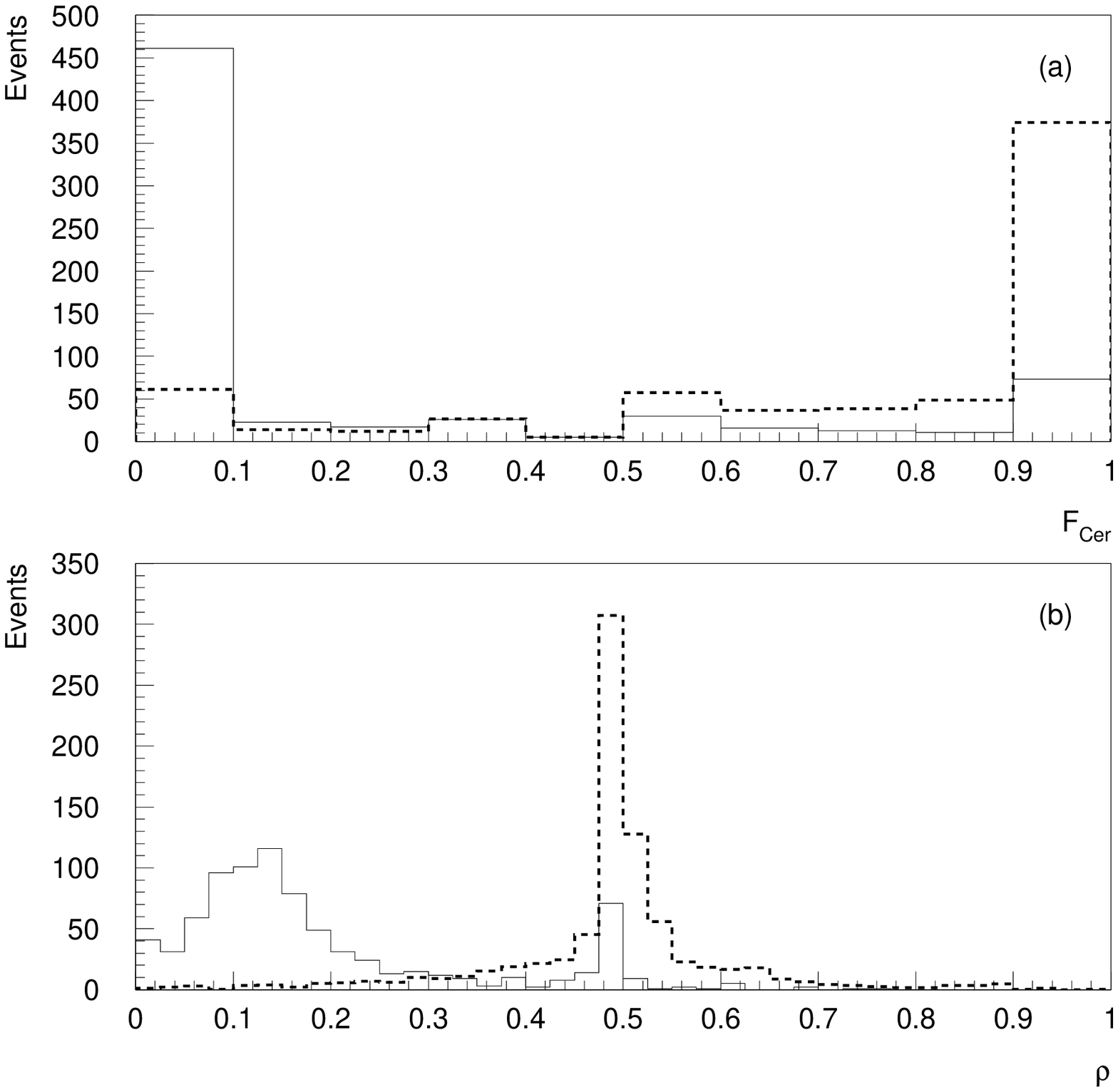}
}
\end{center}
\vspace{5mm}
\caption{
Distributions of (a) the $F_{Cer}$ variable of analysis A and (b) the $\rho$ 
variable of analysis B after all other selections have been applied. 
The solid histograms are DIF data (beam on + off), and the dashed histograms 
are for DIF-MC electrons, normalized to the same area.
}
\label{fig:fcerrho_difmc}
\end{figure}
\newpage
%
\begin{figure}[p]
\vspace*{\fill}
\begin{center}
\mbox{
\epsfxsize=16.0cm \epsfbox[20  20 550 550]{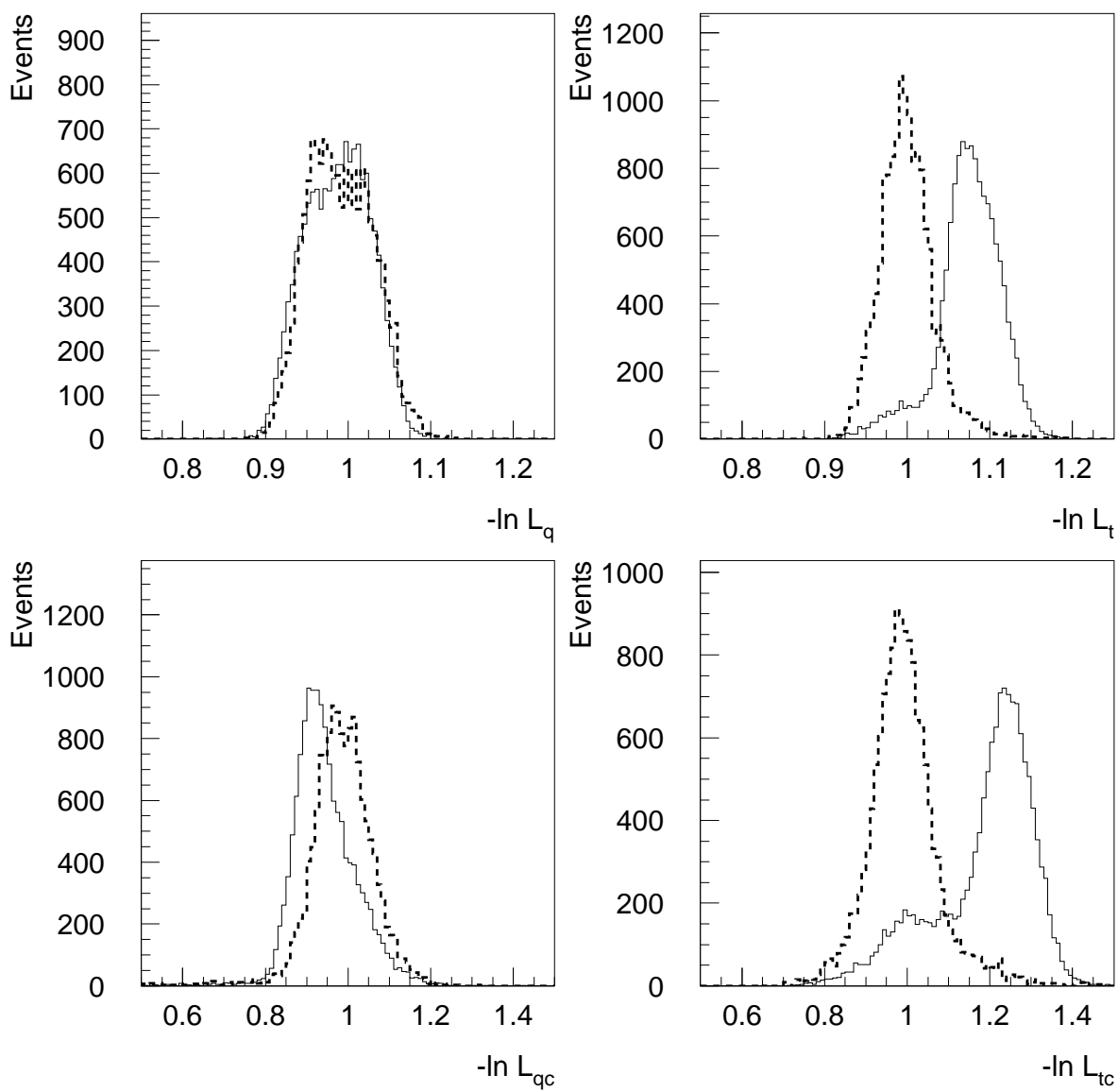}
}
\end{center}
\vspace{5mm}
\caption{
Charge and time negative log-likelihoods for all DIF data (beam on+off) for 
analysis A. 
Superimposed are the same distributions for the DIF-MC electron data (dashed), 
normalized to the same area.
}
\label{fig:pl_ll_dif_he_p}
\end{figure}
\newpage
%
\begin{figure}[p]
\vspace*{\fill}
\begin{center}
\mbox{
\epsfxsize=16.0cm \epsfbox[20  20 550 550]{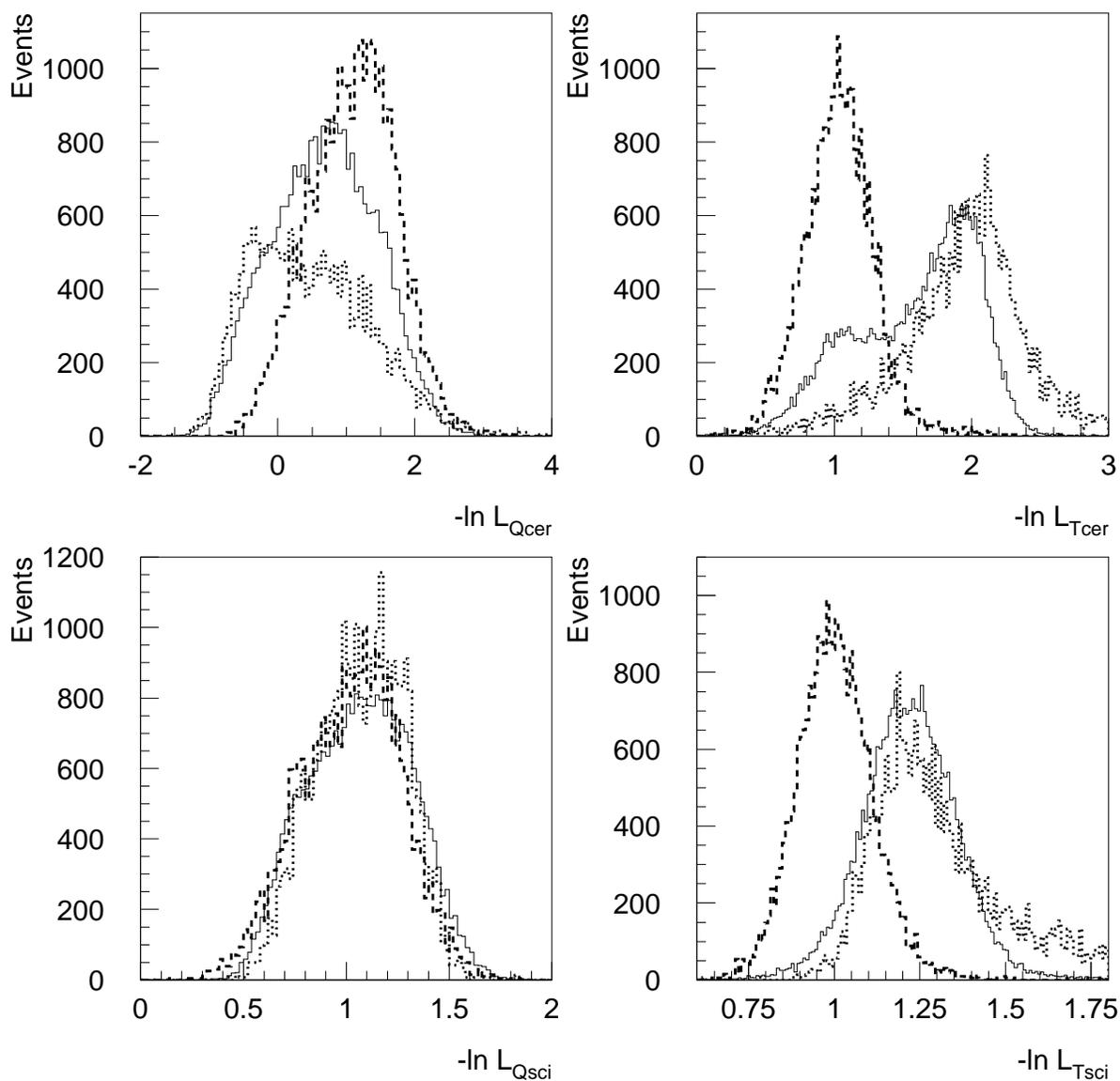}
}
\end{center}
\vspace{5mm}
\caption{
Charge and time negative log-likelihoods for all DIF data (beam on+off) for 
analysis B. 
Superimposed are the same distributions for the DIF-MC electron data (dashed) 
and for cosmic-ray neutrons (dotted), normalized to the same area.
}
\label{fig:llq_difall}
\end{figure}
\newpage
%
\begin{figure}[p]
\vspace*{\fill}
\begin{center}
\mbox{
\epsfxsize=16.0cm \epsfbox[20 270 550 550]{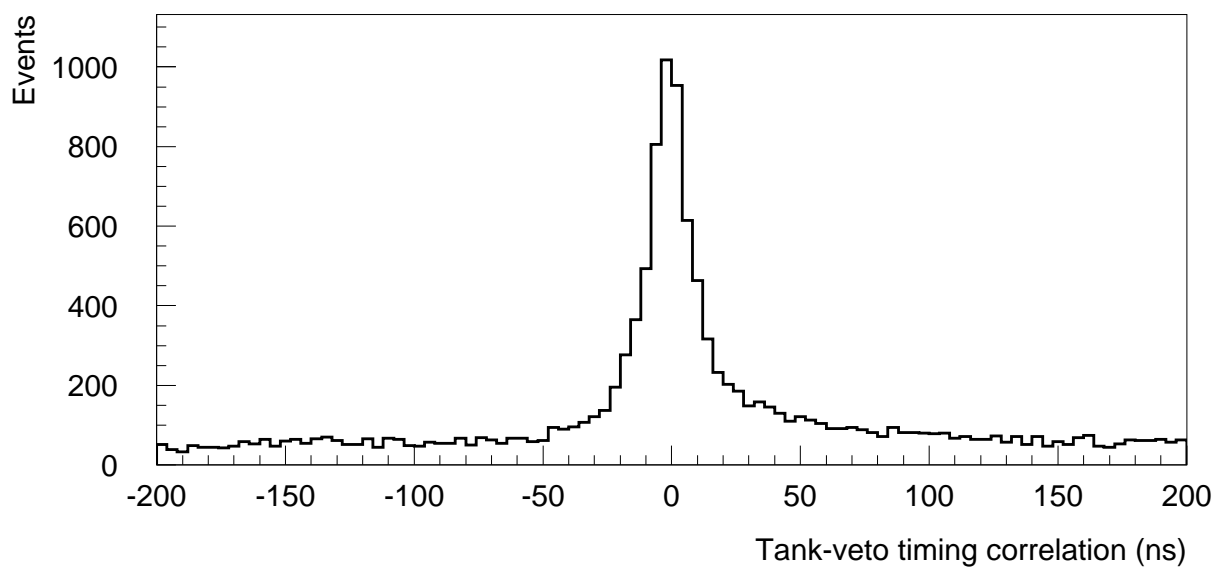}
}
\end{center}
\vspace{5mm}
\caption{
Tank-veto timing correlations for all DIF data (beam on+off). 
The selected events are required to satisfy $|t_{veto}| > 50$ ns in analysis A 
and $|t_{veto}| > 70$ ns in analysis B.
}
\label{fig:tv_timing}
\end{figure}
\newpage
%
\begin{figure}[p]
\vspace*{\fill}
\begin{center}
\mbox{
\epsfxsize=16.0cm \epsfbox[20  20 550 550]{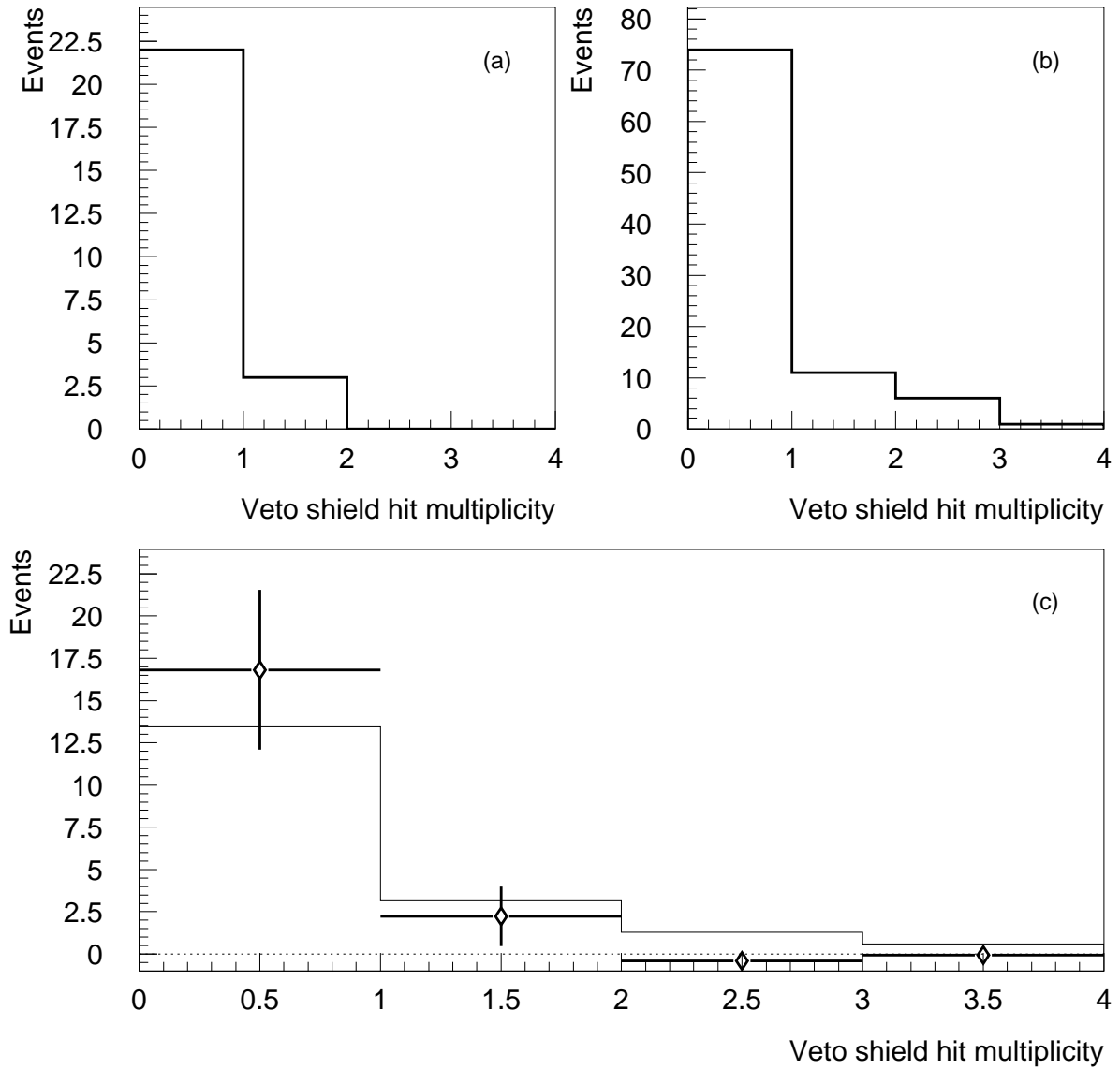}
}
\end{center}
\vspace{5mm}
\caption{
Veto shield hit multiplicity distribution for the events in the final DIF 
sample of analysis B for the (a) beam-on, (b) beam-off and (c) beam-excess 
events. 
The solid histogram in (c) is the expected distribution from laser calibration 
events, which is due only to accidental hits in the veto system.
}
\label{fig:vetoh_dif}
\end{figure}
\newpage
%
\begin{figure}[p]
\vspace*{\fill}
\begin{center}
\mbox{
\epsfxsize=16.0cm \epsfbox[20  20 550 550]{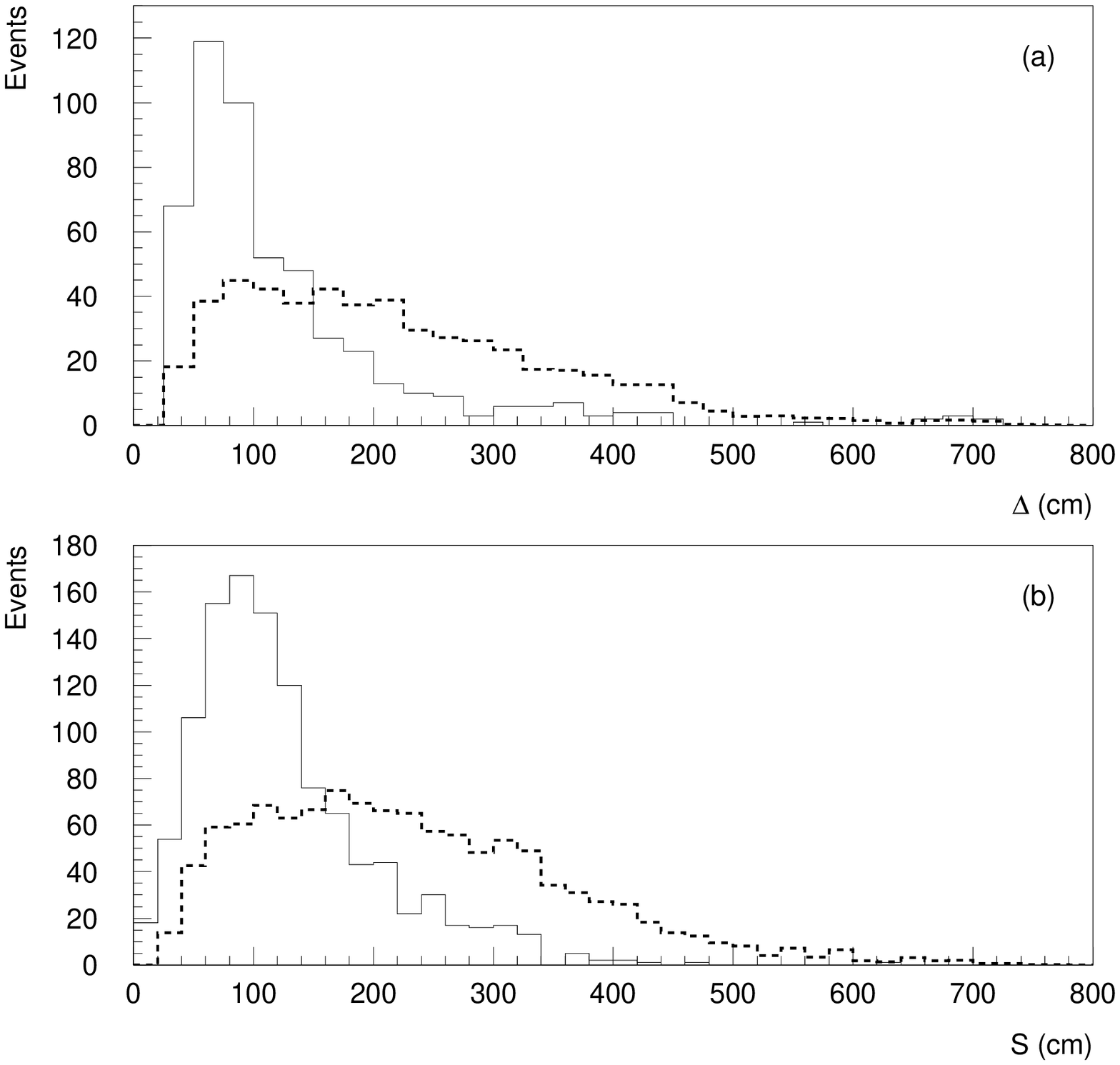}
}
\end{center}
\vspace{5mm}
\caption{
Distributions of (a) the $\Delta$ variable of analysis A and (b) the $S$ 
variable of analysis B after all other selections have been applied. 
The solid histograms are DIF data (beam on + off), and the dashed histograms 
are for DIF-MC electrons, normalized to the same area.
}
\label{fig:deltas_difmc}
\end{figure}
\newpage
%
\begin{figure}[p]
\vspace*{\fill}
\begin{center}
\mbox{
\epsfxsize=16.0cm \epsfbox[20  20 550 550]{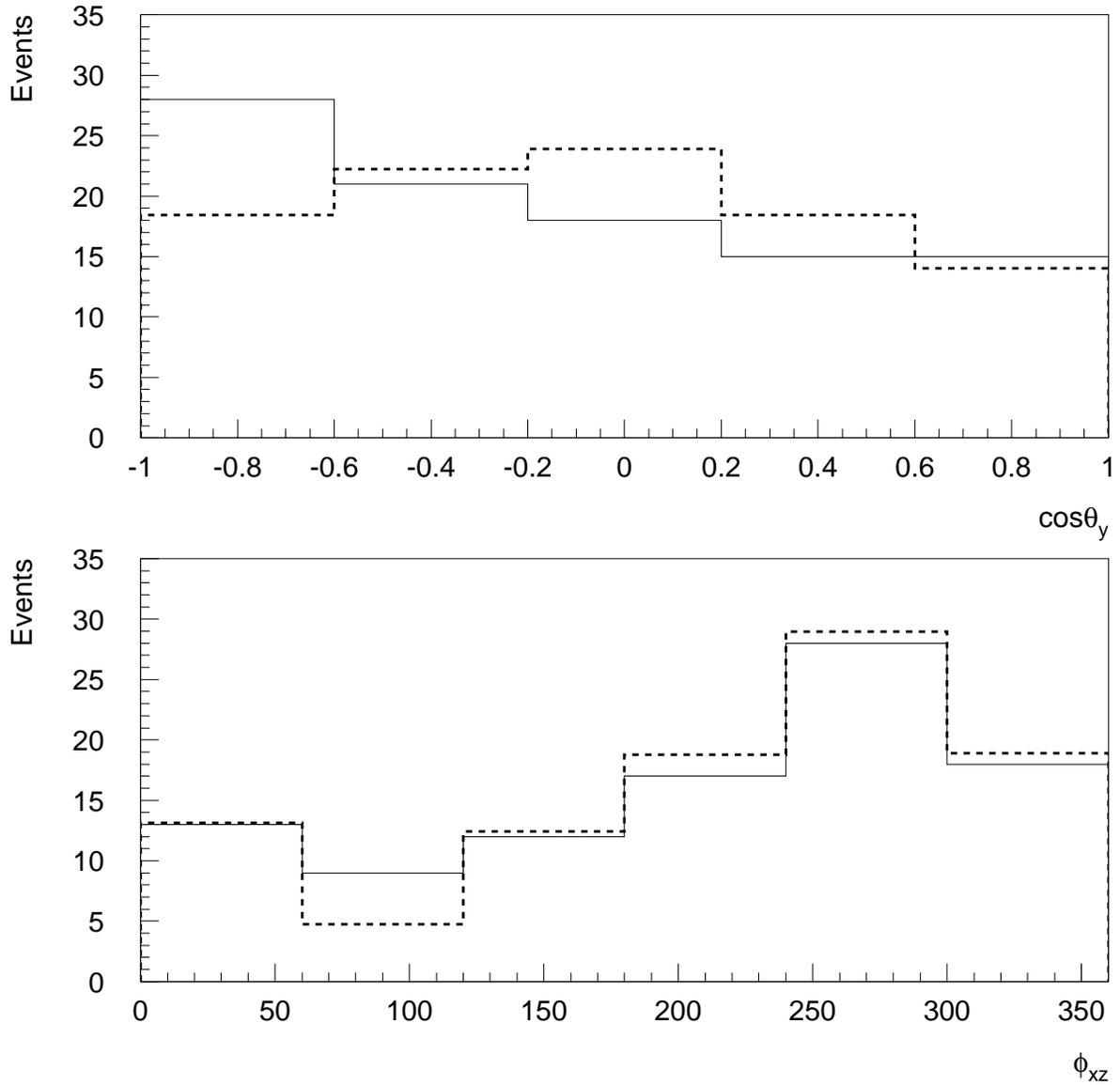}
}
\end{center}
\vspace{5mm}
\caption{
The distribution of event direction in the tank polar coordinate system 
defined by $(\cos\theta_y,\phi_{xz})$. 
The solid histograms show the beam-off DIF data, and the dashed histograms show 
the distributions for DIF-MC electrons (analysis A).
}
\label{fig:cosy_phi}
\end{figure}
\newpage
%
\begin{figure}[p]
\begin{center}
\mbox{
\epsfxsize=16.0cm \epsfbox[20 270 550 550]{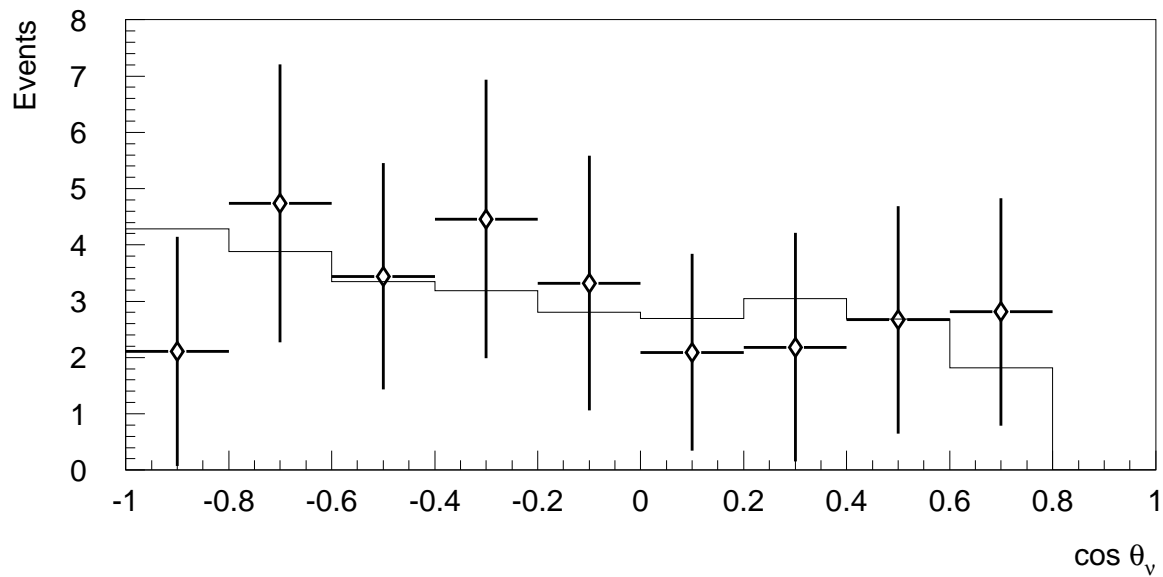}
}
\end{center}
\vspace{5mm}
\caption{
The $\cos \theta_\nu$ distribution, the cosine of the angle between the 
reconstructed event direction and that of the incident neutrino, for the final 
beam-excess DIF data events and that expected from the DIF-MC simulation 
(solid histogram).
}
\label{fig:dif_uzxcs_or}
\end{figure}
%
\begin{figure}[p]
\begin{center}
\mbox{
\epsfxsize=16.0cm \epsfbox[20 270 550 550]{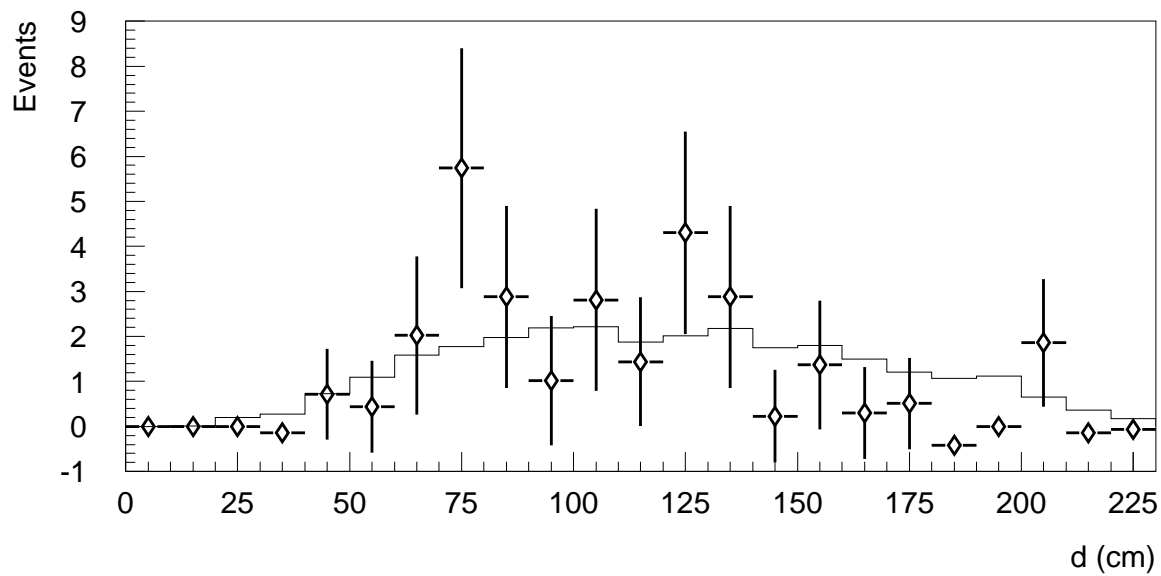}
}
\end{center}
\vspace{5mm}
\caption{
The reconstructed vertex to PMT surface distance distribution for the final 
beam-excess DIF events and that expected from the DIF-MC simulation (solid 
histogram).
}
\label{fig:dif_distxcs_or}
\end{figure}
\newpage
%
\begin{figure}[p]
\vspace*{\fill}
\begin{center}
\mbox{
\epsfxsize=16.0cm \epsfbox[20  20 550 550]{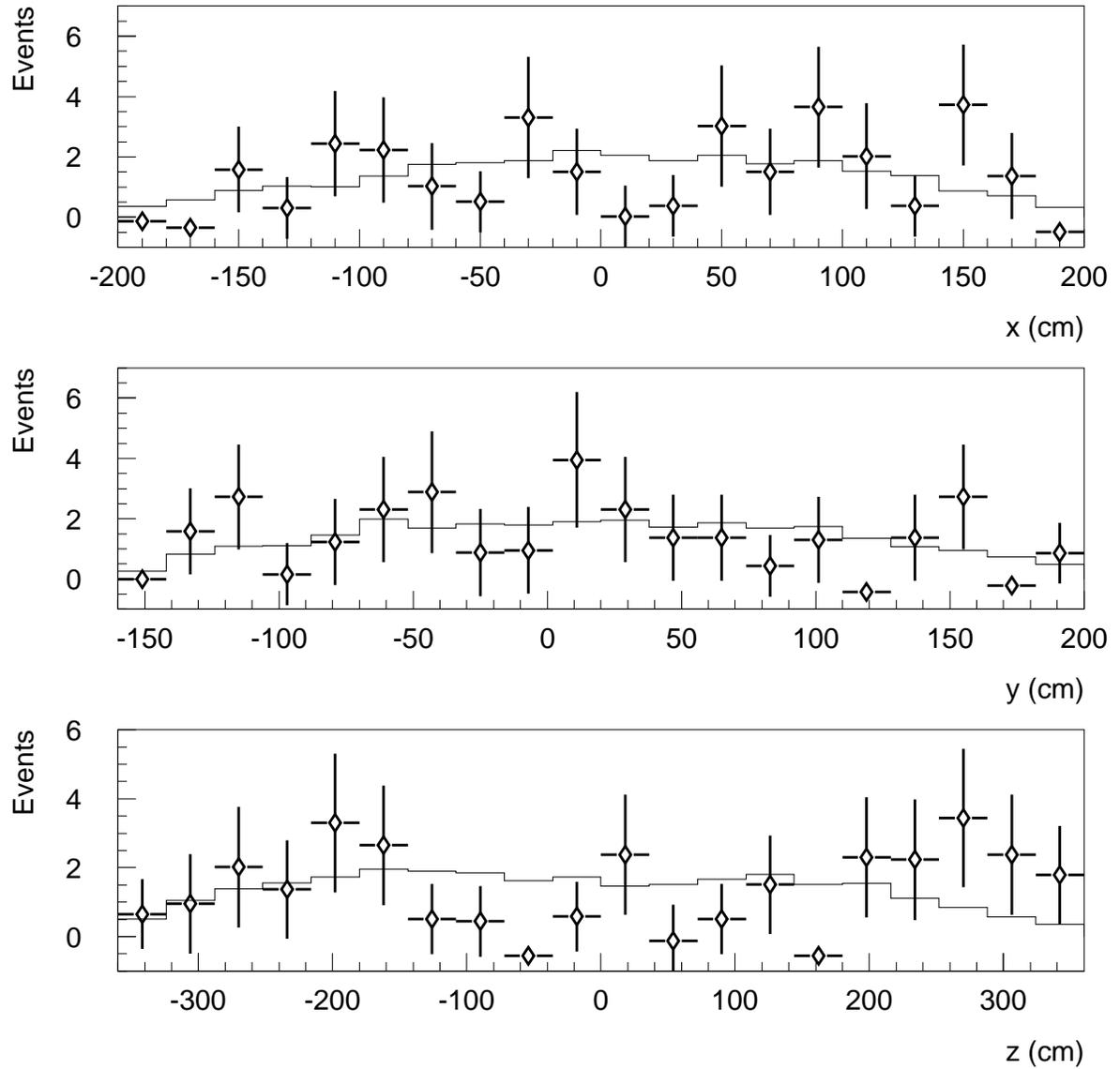}
}
\end{center}
\vspace{5mm}
\caption{
The $x$, $y$ and $z$ distributions for the final beam-excess DIF events and 
those expected from the DIF-MC simulation.
}
\label{fig:dif_xyzxcs_or}
\end{figure}
\newpage
%
\begin{figure}[p]
\vspace*{\fill}
\begin{center}
\mbox{
\epsfxsize=16.0cm \epsfbox[20  20 550 550]{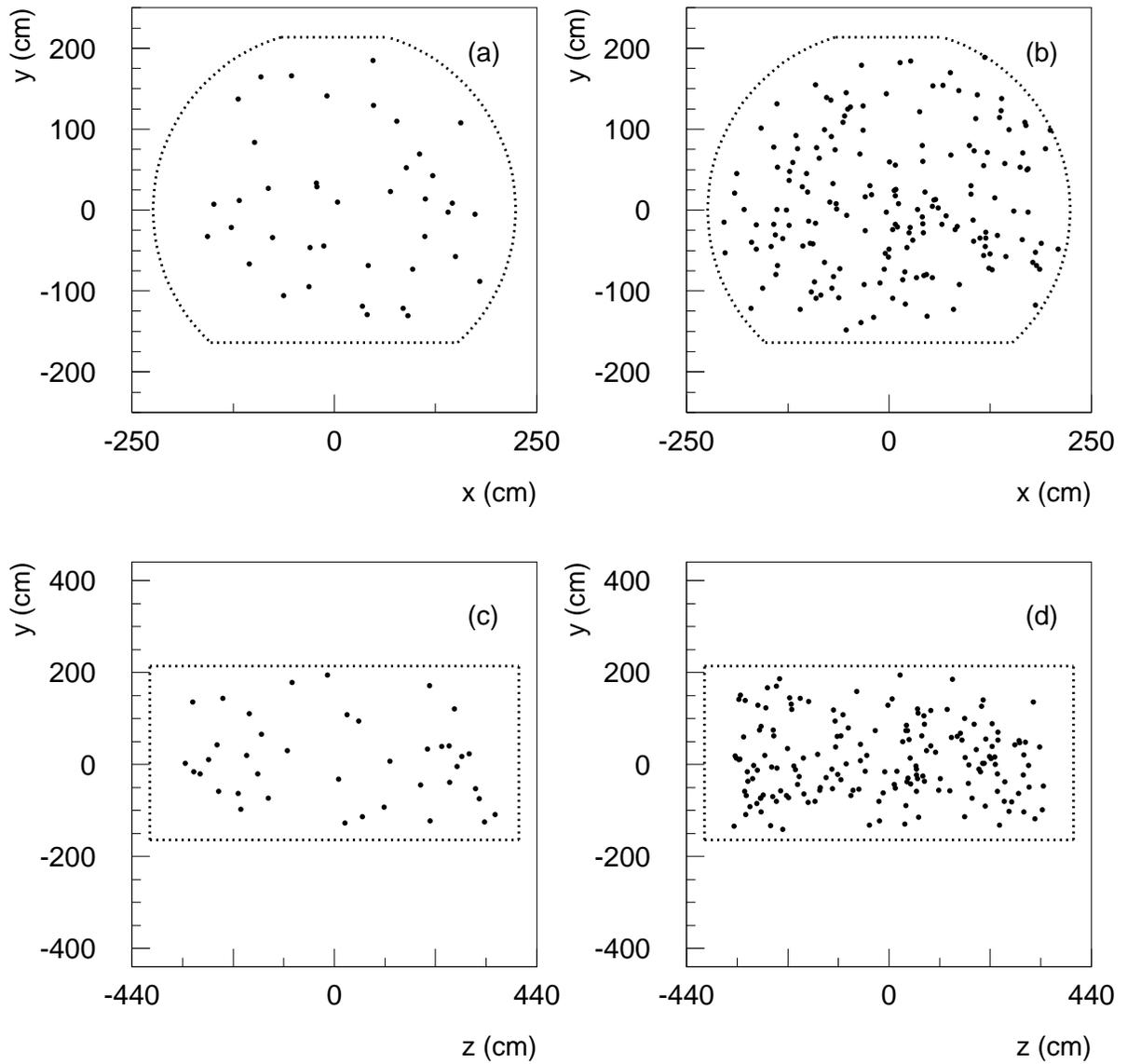}
}
\end{center}
\vspace{5mm}
\caption{
Spatial distributions of the electron events of the final DIF sample in the 
$x-y$ and $y-z$ planes for (a) and (c) the 40 beam-on events and (b) and (d) 
the 175 beam-off events, respectively. 
The dotted contours outline the d $>$ 35 cm fiducial volume.
}
\label{fig:dif_xyyzonoff_or}
\end{figure}
\newpage
%
\begin{figure}[p]
\vspace*{\fill}
\begin{center}
\mbox{
\epsfxsize=15.0cm \epsfbox[20 270 550 550]{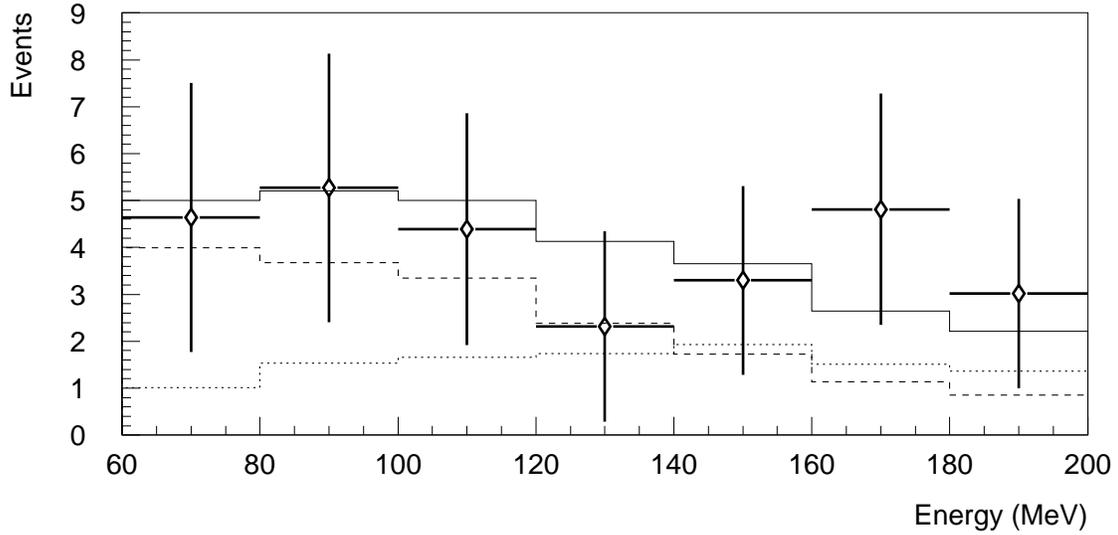}
}
\end{center}
\vspace{3mm}
\caption{
The energy distribution for the final beam-excess DIF events. 
The expectation for backgrounds (dotted histogram), the oscillation signal for 
large values of $\Delta m^2$ (dashed histogram), and the sum of the two (solid 
histogram) are shown also.
}
\label{fig:dif_excs_or}
\end{figure}
%
\begin{figure}[p]
\begin{center}
\mbox{
\epsfxsize=15.0cm \epsfbox[20 270 550 550]{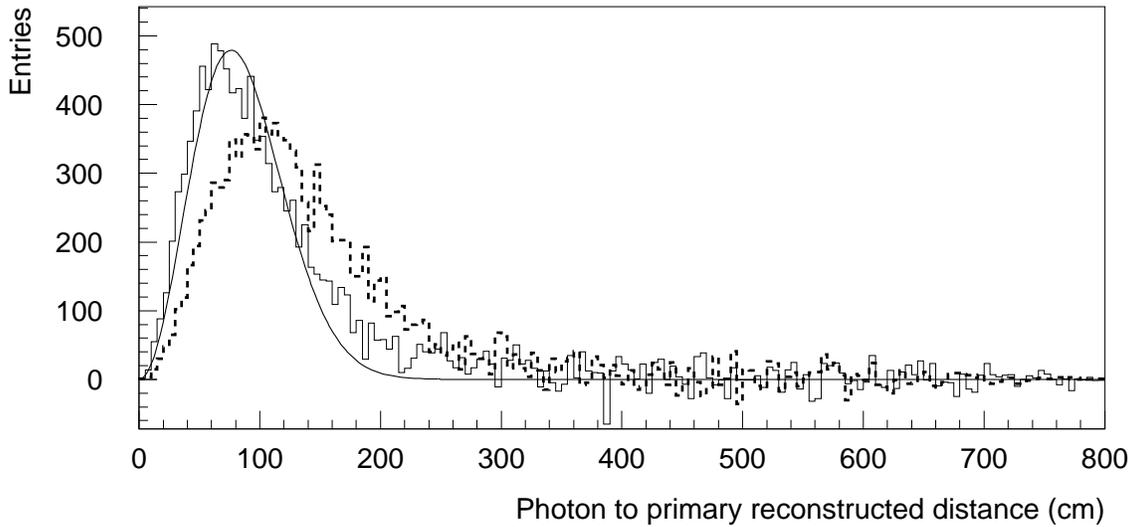}
}
\end{center}
\vspace{3mm}
\caption{
Photon to primary reconstructed distance for the old (dashed histogram) and new 
reconstruction algorithms (solid histogram) for correlated photons. 
The average distance is decreased from 76 cm to 54 cm. 
The superimposed fit is to $f(dr)$ = $C\,dr^2\,\exp[-dr^2/(2\sigma^2)]$.
}
\label{fig:gdr_oldnew}
\end{figure}
\newpage
%
\begin{figure}[p]
\vspace*{\fill}
\begin{center}
\mbox{
\epsfxsize=16.0cm \epsfbox[20  20 550 550]{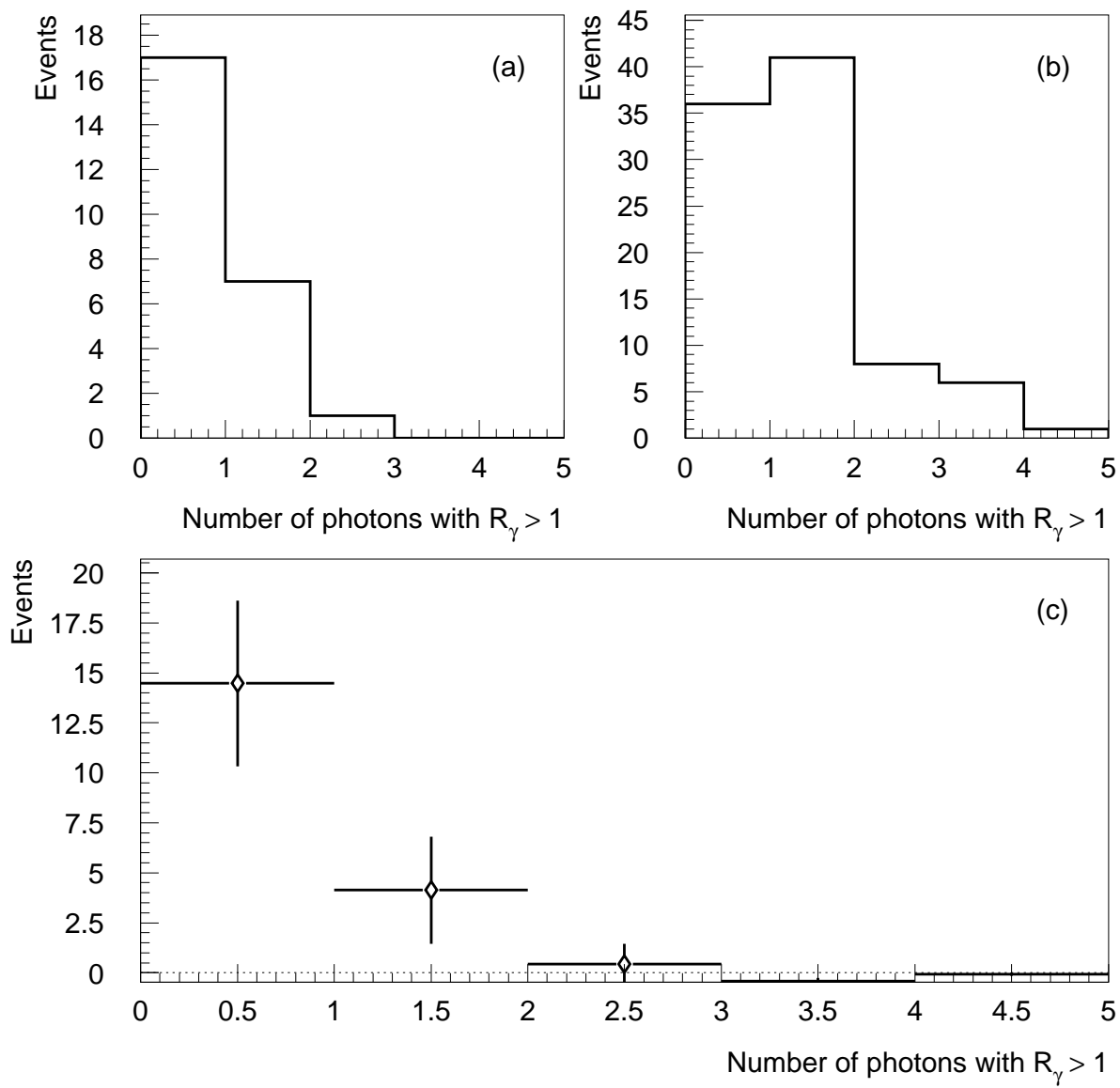}
}
\end{center}
\vspace{5mm}
\caption{
Distribution of the number of photons with $R_{\gamma} > 1$ in the final DIF 
sample for the (a) beam-on, (b) beam-off and (c) beam-excess events 
(analysis B).
}
\label{fig:corgammas_dif}
\end{figure}
\newpage
%
\begin{figure}[p]
\vspace*{\fill}
\begin{center}
\mbox{
\epsfxsize=16.0cm \epsfbox[20  20 550 550]{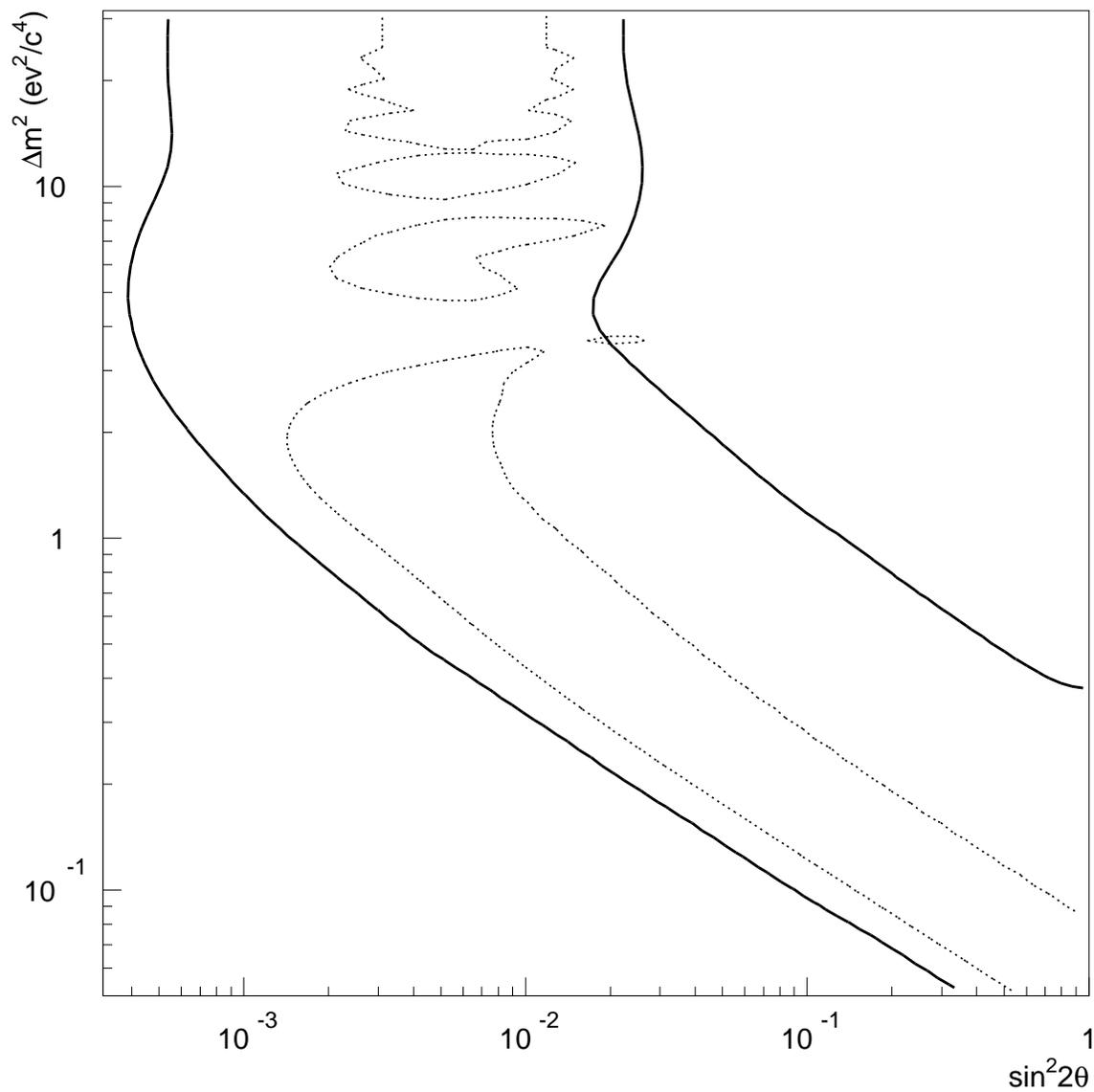}
}
\end{center}
\vspace{5mm}
\caption{
The 95\% confidence region for the DIF $\nu_\mu \to \nu_e$ along with the 
favored regions for the LSND DAR measurement for $\bar\nu_\mu \to \bar\nu_e$. 
}
\label{fig:confidence}
\end{figure}
%
%
%

%
%
\end{document}